\documentclass[12pt,preprint]{aastex}

\usepackage{natbib}
\usepackage{graphicx}
\usepackage{url}

\newcommand{\Msun}{\ensuremath{\mathrm{M_{\Sun}}}} 
\newcommand{\yr}{\textrm{yr}} 
\newcommand{\ms}{\textrm{ms}} 
\newcommand{\MHz}{\textrm{MHz}} 
\newcommand{\GHz}{\textrm{GHz}} 
\newcommand{\us}{\ensuremath\mathrm{\mu s}} 
\newcommand{\sn}{\textrm{S/N}} 
\newcommand{\rmsub}[1]{\ensuremath{_{\mathrm{#1}}}}
\newcommand{\s}{\textrm{s}} 
\newcommand{\K}{\textrm{K}} 
\newcommand{\Jy}{\textrm{Jy}} 
\newcommand{\Hz}{\textrm{Hz}} 
\newcommand{\pmsq}{\ensuremath{\mathrm{m^{-2}}}} 
\newcommand{\hr}{\textrm{hr}} 
\newcommand{\Rsun}{\ensuremath{\mathrm{R_{\Sun}}}} 
\newcommand{\dm}{\textrm{DM}} 
\newcommand{\km}{\textrm{km}} 
\newcommand{\pc}{\textrm{pc}} 
\newcommand{\Gyr}{\textrm{Gyr}} 
\newcommand{\ps}{\ensuremath{\mathrm{s^{-1}}}} 
\newcommand{\Lsun}{\ensuremath{\mathrm{L_{\Sun}}}} 
\newcommand{\pyr}{\ensuremath{\mathrm{yr^{-1}}}} 
\newcommand{\gauss}{\textrm{G}} 
\newcommand{\kpc}{\textrm{kpc}} 
\newcommand{\dmu}{\ensuremath{\mathrm{pc\; cm^{-3}}}}
\newcommand{\erg}{\textrm{erg}} 
\newcommand{\uJy}{\ensuremath{\mathrm{\mu Jy}}} 

\title{The Timing of Nine Globular Cluster Pulsars}

\author{Ryan S.\ Lynch\altaffilmark{1,2}, Paulo C.\ C.\
  Freire\altaffilmark{3}, Scott M.\ Ransom\altaffilmark{4}, and Bryan
  A.\ Jacoby\altaffilmark{5}}
\altaffiltext{1}{Physics Department, McGill University, 3600 Rue
  University, Montr\'{e}al, QC, Canada H3A 2T8,
  \texttt{rlynch@physics.mcgill.ca}}
\altaffiltext{2}{Department of Astronomy, University of Virginia,
  P.O.\ Box 400325, Charlottesville, VA 22904-4325}
\altaffiltext{3}{Max-Planck-Institut f\"{u}r Radioastronomie, Auf dem
  H\"{u}gel 69, D-53121 Bonn, Germany,
  \texttt{pfreire@mpifr-bonn.mpg.de}}
\altaffiltext{4}{National Radio Astronomy Observatory, 520 Edgemont
  Road, Charlottesville, VA 22903-4325, \texttt{sransom@nrao.edu}}
\altaffiltext{5}{Affiliated with The Aerospace Corporation, 15049
  Conference Center Drive, Chantilly, VA 20151-3824,
  \texttt{bryan.jacoby@gmail.com}}

\shorttitle{Timing Nine GC Pulsars}
\shortauthors{Lynch et al.}

\begin{document}

\setcounter{footnote}{5}

\begin{abstract}

  We have used the Robert C.\ Byrd Green Bank Telescope to time nine
  previously known pulsars without published timing solutions in the
  globular clusters M62, NGC~6544, and NGC~6624.  We have full timing
  solutions that measure the spin, astrometric, and (where applicable)
  binary parameters for six of these pulsars.  The remaining three
  pulsars (reported here for the first time) were not detected enough
  to establish solutions.  We also report our timing solutions for
  five pulsars with previously published solutions, and find good
  agreement with past authors, except for PSR J1701$-$3006B in M62.
  Gas in this system is probably responsible for the discrepancy in
  orbital parameters, and we have been able to measure a change in the
  orbital period over the course of our observations.  Among the
  pulsars with new solutions we find several binary pulsars with very
  low mass companions (members of the so-called ``black widow'' class)
  and we are able to place constraints on the mass-to-light ratio in
  two clusters.  We confirm that one of the pulsars in NGC~6624 is
  indeed a member of the rare class of non-recycled pulsars found in
  globular clusters.  We also have measured the orbital precession and
  Shapiro delay for a relativistic binary in NGC~6544.  If we assume
  that the orbital precession can be described entirely by general
  relativity, which is likely, we are able to measure the total system
  mass ($2.57190(73)\; \Msun$) and companion mass ($1.2064(20)\;
  \Msun$), from which we derive the orbital inclination ($\sin{i} =
  0.9956(14)$) and the pulsar mass ($1.3655(21)\; \Msun$), the most
  precise such measurement ever obtained for a millisecond pulsar. The
  companion is the most massive known around a fully recycled pulsar.
\end{abstract}


\maketitle

\section{Introduction \label{sec:intro}}

Millisecond pulsars (MSPs) form by ``recycling'' a dormant neutron
star through the accretion of mass and angular momentum from a binary
companion \citep{acr+82}.  This leads to a very rapidly rotating and
stable pulsar with a relatively low magnetic field ($\sim 10^9\;
\gauss$) and long lifetime ($\gtrsim 10^9\; \yr$).  MSPs form
naturally in the dense environments of globular clusters (GCs), thanks
to frequent exchange interactions that may lead to the formation of
mass transferring binaries \citep{cr05}.  Sensitive searches of GCs
have uncovered 144 pulsars in 28 clusters\footnote{See
  \url{http://www.naic.edu/~pfreire/GCpsr.html} for an up-to-date
  list.}, and the vast majority of these are recycled MSPs.  Indeed,
nearly half of all MSPs have been discovered in GCs\footnote{There are
  160 field pulsars with period $P < 20\; \ms$ in the ATNF catalog,
  and 131 GC pulsars that meet the same criteria (see
  \url{http://www.atnf.csiro.au/people/pulsar/psrcat/}).}.

The same interactions that form MSPs so efficiently in clusters also
lead to many exotic systems that are rarely seen in the disk of the
Galaxy.  These include the fastest spinning MSP \citep{hrs+06}, highly
eccentric binaries \citep{rhs+05,frg07}, massive neutron stars
\citep{frb+08}, pulsar-main sequence binaries \citep{dpm+01}, and many
``black widow'' systems \citep{kbr+05}.  These discoveries demonstrate
the huge scientific payoff that can come from the discovery of unique
pulsars thanks to two factors---the extreme nature of neutron stars,
which opens windows on otherwise inaccessible realms of physics, and
the extraordinary clock-like nature of MSPs.  The bedrock of pulsar
astronomy is \emph{timing}, the process of creating a model that
unambiguously accounts for every rotation of the pulsar, thus probing
the pulsar and its environment.  For MSPs, the arrival time of a pulse
can typically be measured to within a few microseconds or better
\citep[e.g.][]{vbv+08}, which enables very precise timing models.
Timing GC pulsars leads to unique challenges, though.  Because GCs are
usually at a distance of several kiloparsecs, the flux density of
their constituent MSPs is usually very low, necessitating very long
integration times.  Another consequence of cluster distances is very
high dispersion measures, which necessitate moving to higher observing
frequencies (e.g. $2\; \GHz$) where pulsars tend to be weaker.
Luckily, since many clusters contain several MSPs that can each be
observed during a single observation, the required observing time is
well spent.

In addition to recycled MSPs, a small population of slow, non-recycled
pulsars have also been observed in a handful of clusters
\citep{lmd96,blt+11}.  Their presence is something of a mystery,
because they resemble in every way the ``normal'' pulsars that are so
numerous in the Galactic disk, but which have lifetimes $\sim
10^7$--$10^8\; \yr$ and are thought to form through core collapse
supernovae of massive stars.  However, GCs are old stellar systems
with typical ages $10^{10}\; \yr$ \citep{cgcf00}, so
all stars massive enough to form pulsars should have died some
$10^{10}\; \yr$ ago.  As such, the core collapse of massive stars
cannot be the avenue through which non-recycled pulsars in GCs have
formed.  The most popular alternative formation scenario involves the
collapse of a massive white dwarf via electron capture
\citep{nom84,nom87}, though the details of these so-called electron
capture supernovae are not well understood.

In this paper, we present new timing solutions for nine pulsars spread
across three GCs---M62, NGC~6544, and NGC~6624.  Each of these
clusters contains multiple MSPs and NGC~6624 contains two non-recycled
pulsars.  All of these pulsars were discovered elsewhere, but full,
phase-connected timing solutions have only been published for five of
them (see Table \ref{table:psrs} for a summary and list of
references).  We also present our solutions for these five.  We began
a timing campaign using the Robert C.\ Byrd Green Bank Telescope (GBT)
for the nine pulsars without full solutions, starting in 2009
February.  We have managed to obtain full solutions that include
measurement of the first period derivative for six of these pulsars.
A combination of low flux densities and an irreversible data
processing error conspired to prevent us from obtaining full solutions
for the two remaining MSPs, but we are in a position to comment on
them in further detail.  Our most exciting result is the discovery of
a massive companion orbiting a fully recycled MSP.  In \S\ref{sec:obs}
we describe our observations and in \S\ref{sec:timing} outline our
method for timing the pulsars and further data analysis.  A discussion
of individual systems can be found in \S\ref{sec:psrs}, and we provide
a summary in \S\ref{sec:conc}.

\section{Observing Scheme \label{sec:obs}}

Observations were carried out with the GBT, observing at a frequency
of $2.0\; \GHz$ using $800\; \MHz$ of bandwidth, although persistent
radio frequency interference (RFI) reduced the usable bandwidth to
$\sim 600\; \MHz$.  This frequency was chosen to overcome the
deleterious effects of dispersive smearing caused by free electrons in
the ISM, and has been used successfully by our group before.  Data
were recorded in the
\texttt{PSRFITS}\footnote{\url{http://www.atnf.csiro.au/research/pulsar/psrfits/index.html}}
\citep{hvm04} format using the Green Bank Ultimate Pulsar Processor
(GUPPI) \citep{drd+08} with a $64\; \us$ sampling time and 2048
frequency channels across the entire bandwidth.  We typically observed
NGC~6544 for 30 minutes, and M62 and NGC~6624 for 45--60 minutes each,
although the exact integration times varied between observations.  In
general, data were relatively free of RFI, but when necessary, we used
the RFI excision tools in the
\texttt{PRESTO}\footnote{\url{http://www.cv.nrao.edu/~sransom/presto/}}
software suite to mask out contaminated portions of the data.

We also used archival GBT data taken in 2004.  These data were
collected using the GBT Pulsar Spigot \citep{kel+05} at either $820\;
\MHz$ (with $50\; \MHz$ of bandwidth) or at $2\; \GHz$ (with the same
bandwidth as above).  Most of these observations used 2048 frequency
channels and $40.96\; \us$ sampling, although some used 1024 channels
and $81.92\; \us$ sampling.  All of these data were analyzed using
\texttt{PRESTO}.

The pulsars in NGC~6544 were observed again in 2011 as part of a
campaign to measure Shapiro delay in PSR J1807$-$2500B (see
\ref{sec:NGC6544B}).  These data were collected using GUPPI in a
coherent de-dispersion search mode (i.e. a filterbank where each
channel was coherently de-dispersed) at $1.4\; \GHz$ using $800\;
\MHz$ of bandwidth.  For these observations we used 512 frequency
channels and $10.24\; \us$ sampling.

\section{Timing and Data Analysis \label{sec:timing}}

Data were processed using a combination of \texttt{PRESTO} and
\texttt{PSRCHIVE}\footnote{\url{http://psrchive.sourceforge.net/}}.
All of the folded pulse profiles were phase-aligned and summed to
create high signal-to-noise (\sn) profiles.  The profiles were fit
with one or more Gaussians, from which standard pulse templates were
made and used to obtain pulse times of arrival (TOAs) via
cross-correlation in the Fourier domain.  We typically obtained three
to six TOAs per pulsar per observation depending on the \sn\ of that
observation, though some pulsars (such as NGC~6544A) were sufficiently
bright that many precise TOAs could be measured.  Timing solutions
were constructed by performing a weighted fit to the data using
\texttt{TEMPO2}\footnote{\url{http://sourceforge.net/projects/tempo2/}}
\citep{ehm06} with the DE405 Solar System ephemeris and TT(BIPM2011) time
standard.  All values are reported in Barycentric Dynamical Time.  It
is not uncommon for timing models to have a reduced $\chi^2 > 1$
($\chi^2\rmsub{red}$) even after all parameters have been well
measured.  When no systematic trends are present in the data, it is
assumed that this value of $\chi^2\rmsub{red}$ is due to an
underestimate of the error on individual TOAs.  As is common practice,
we deal with this situation by multiplying TOA errors by a small
constant error factor so that $\chi^2\rmsub{red} = 1$.

\subsection{Corrections for Cluster Acceleration \label{sec:cluster_a}}

MSPs have a small intrinsic rate of spin-down ($\dot{P}\rmsub{int}$)
that is usually heavily contaminated by accelerations within the
potential of the cluster and Galaxy \citep{phi93}.  This makes it
impossible to measure their spin-down related properties (surface
magnetic field, spin-down luminosity, and characteristic age)
directly.  Instead, we calculate the limit
\begin{eqnarray}
\frac{\dot{P}\rmsub{int}}{P} \leq \frac{\dot{P}\rmsub{obs}}{P} +
                                  \frac{a\rmsub{c,max}}{c} + 
                                  \frac{a\rmsub{G}}{c} +
                                  \frac{\mu^2 D}{c}
\label{eqn:pdot}
\end{eqnarray}
where $P$ is the pulsar period, $a\rmsub{c,max}$ and $a\rmsub{G}$ are
the accelerations due to the cluster and Galaxy, respectively, $\mu$
is the proper motion, $D$ is the distance to the cluster, and $c$ is
the speed of light (the last term accounts for the Shklovskii effect
\citep{shk70}).  \citet{phi92,phi93} showed that to within 10\% accuracy,
\begin{eqnarray}
\frac{a\rmsub{c,max}}{c} \approx \frac{3 \sigma_v^2}
                                      {2 c (R\rmsub{c}^2 + R\rmsub{psr}^2)^{1/2}}
\label{eqn:amax}
\end{eqnarray}
for $R\rmsub{psr} < 2 R\rmsub{c}$, where $\sigma_v$ is the central
velocity dispersion, $R\rmsub{c}$ is the core radius, and
$R\rmsub{psr}$ is the projected distance of the pulsar from the center
of the cluster.  Because all three of the clusters studied here are
core collapsed, analytical \citep{fhn+05} and numerical models are not
applicable, which is why we use the approximations of
\citeauthor{phi92}.  These were developed for use with pulsars in M15,
which is also core collapsed.  We use the values of $R\rmsub{c}$ found
in \citet[2010 edition]{har96}.  The following references are used for
$\sigma_v$: \citet[2010 edition]{har96} for M62; \citet{web85} for
NGC~6544; and \citet{vor11} for NGC~6624.  Table \ref{table:gcs} lists
the relevant properties of each cluster.  The Galactic contribution is
calculated under the approximation of a spherically symmetric Galaxy
with a flat rotation curve \citep{phi93} and is
\begin{eqnarray}
  \dot{P}\rmsub{Gal} = -7 \times 10^{-19} \left ( \frac{P}{\s} \right )
  \left ( \cos{b} \cos{\ell} + 
    \frac{\delta - \cos{b} \cos{\ell}}
    {1 + \delta^2 - 2 \delta \cos{b} \cos{\ell}} \right )
\end{eqnarray}
where $b$ and $\ell$ are the Galactic latitude and longitude of the
cluster, $\delta = R_0/D$ and $R_0$ is the Sun's Galactocentric
distance.  However, the cluster term is usually dominant.

\subsection{Flux Calibration and Rotation Measure \label{sec:flux_rm}}

\begin{figure}
\centering
\includegraphics[height=6.0in]{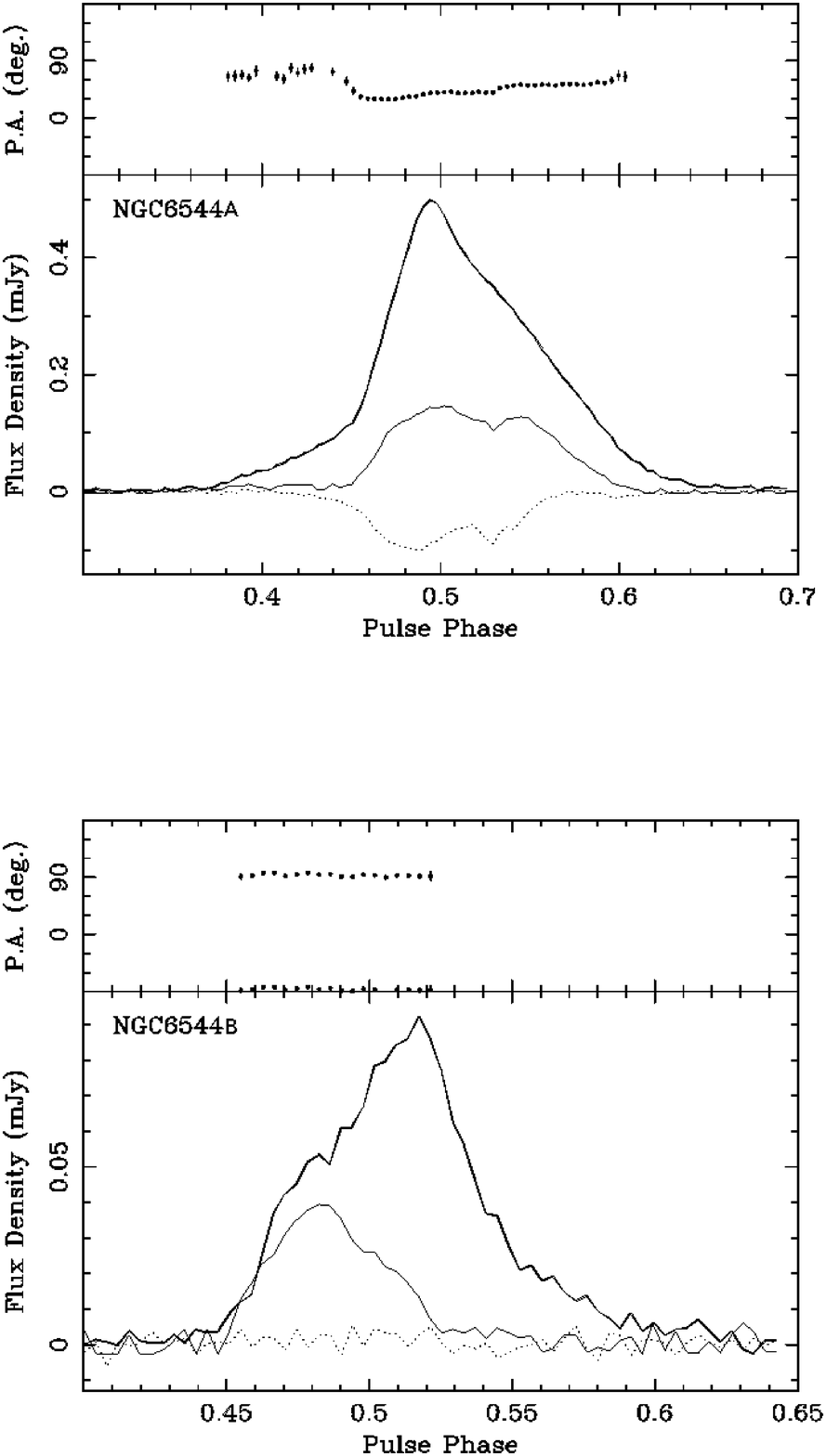}
\caption{Calibrated pulse profiles for NGC~6544A and B showing the
  integrated pulse profile (top solid lines) and the fraction of
  linear (second solid line) and circular (dotted line) polarization.
  The top panels show the polarization angle.  The profiles have been
  rotated by an arbitrary amount.  These data are from our Shapiro
  delay observations for NGC~6544B (see \S\ref{sec:NGC6544B}) and were
  taken at $1.4\; \GHz$ in a coherent de-dispersion mode.  The
  sampling time was $10.24\; \us$ ($0.3\%$ and $0.2\%$ of pulse phase
  for NGC~6544A and B, respectively). \label{fig:full_stokes}}
\end{figure}

Rough mean flux density ($S_{\nu}$) estimates were made by assuming
that the off-pulse RMS noise level was described by the radiometer
equation,
\begin{eqnarray}
  \sigma = \frac{T\rmsub{tot}}
                {G \sqrt{n\rmsub{pol} \Delta \nu\, t\rmsub{obs}}}
\end{eqnarray}
where $T\rmsub{tot}$ is the total system temperature, $G$ is the
telescope gain, $n\rmsub{pol} = 2$ is the number of summed
polarizations, and $\Delta \nu$ is the bandwidth.  For the GBT $2\;
\GHz$ receiver, $G = 1.9\; \K\: \Jy^{-1}$ and $T\rmsub{sys} \approx
23\; \K + T\rmsub{sky}$, where $T\rmsub{sky}$ is the contribution from
the Galactic synchrotron emission.  This was calculated by scaling the
values from \citet{hss+82} with a spectral index of $-2.6$.  The
typical uncertainty in these estimates of $S_{\nu}$ is 10\%--20\%.  We
were able to record full polarization data for one observation per
cluster.  These were calibrated using the $25\; \Hz$ noise diode on
the GBT, and we searched for a significant rotation measure (RM) by
looking for a peak in the polarized flux.  We searched from $-1000$ to
$1000\; \mathrm{rad}\: \pmsq$, but in most cases, the \sn\ was not
sufficient to constrain RM.  The notable exceptions are the two
pulsars in NGC~6544, for which we measure an average RM of $\sim 158\;
\mathrm{rad}\: \pmsq$.  Figure \ref{fig:full_stokes} shows the fully
calibrated, RM corrected profiles of NGC~6544A and B.

\section{Discussion of Individual Systems \label{sec:psrs}}

\begin{figure}
\centering
\includegraphics[height=6.5in]{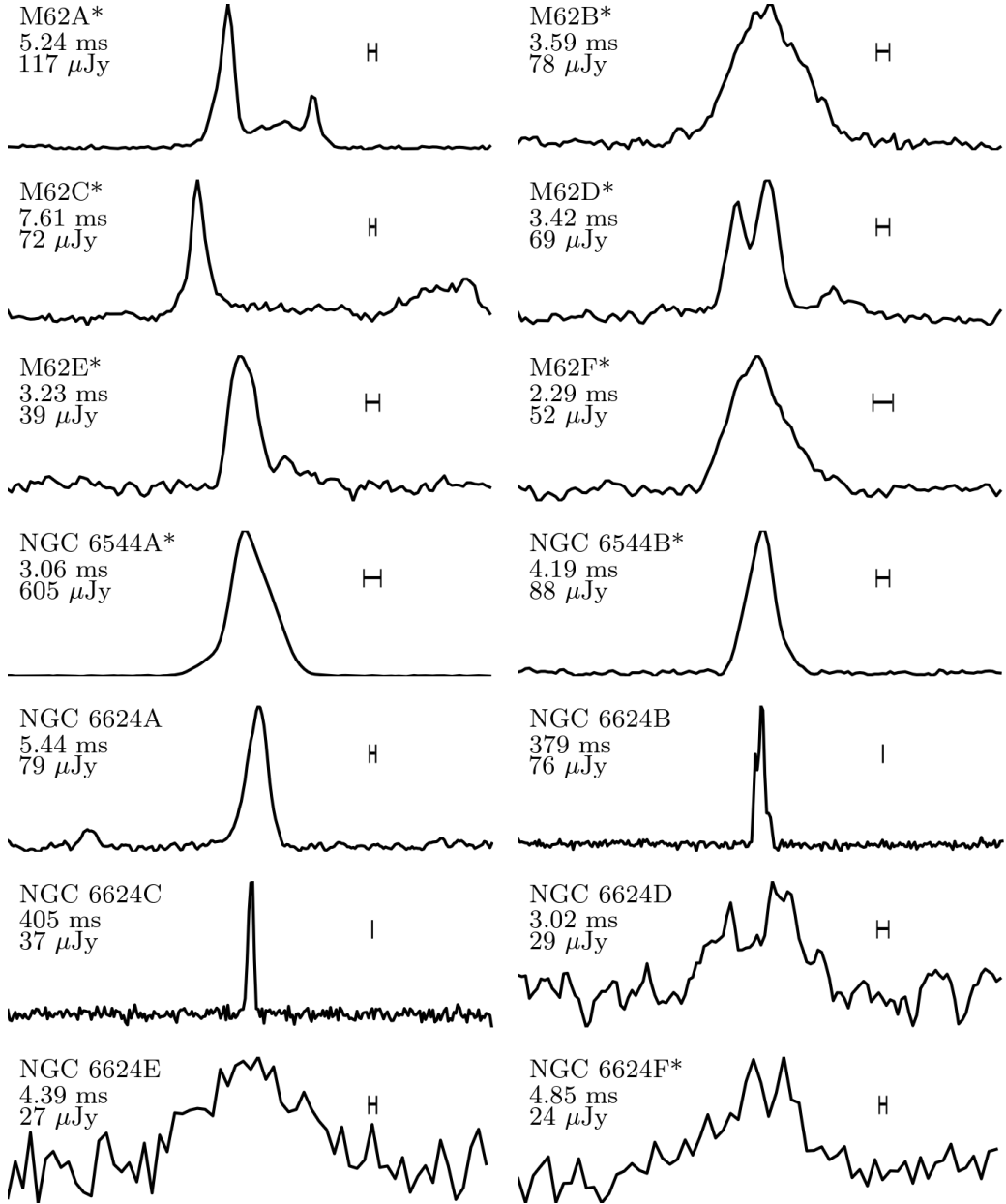}
\caption{Average $2\; \GHz$ pulse profiles for all pulsars in these
  clusters.  Binary pulsars are denoted by an asterisk.  These
  profiles formed the basis of the standards used in our timing
  analysis, and were created by summing all the detections of a given
  pulsar.  In the case of the pulsars in NGC~6624, only data
  de-dispersed at the correct DM was used for making these plots (see
  text for more details).  The bars indicate the contribution of
  dispersive smearing to the profile width. \label{fig:profs}}
\end{figure}

\begin{figure}
\centering
\includegraphics[width=5.75in]{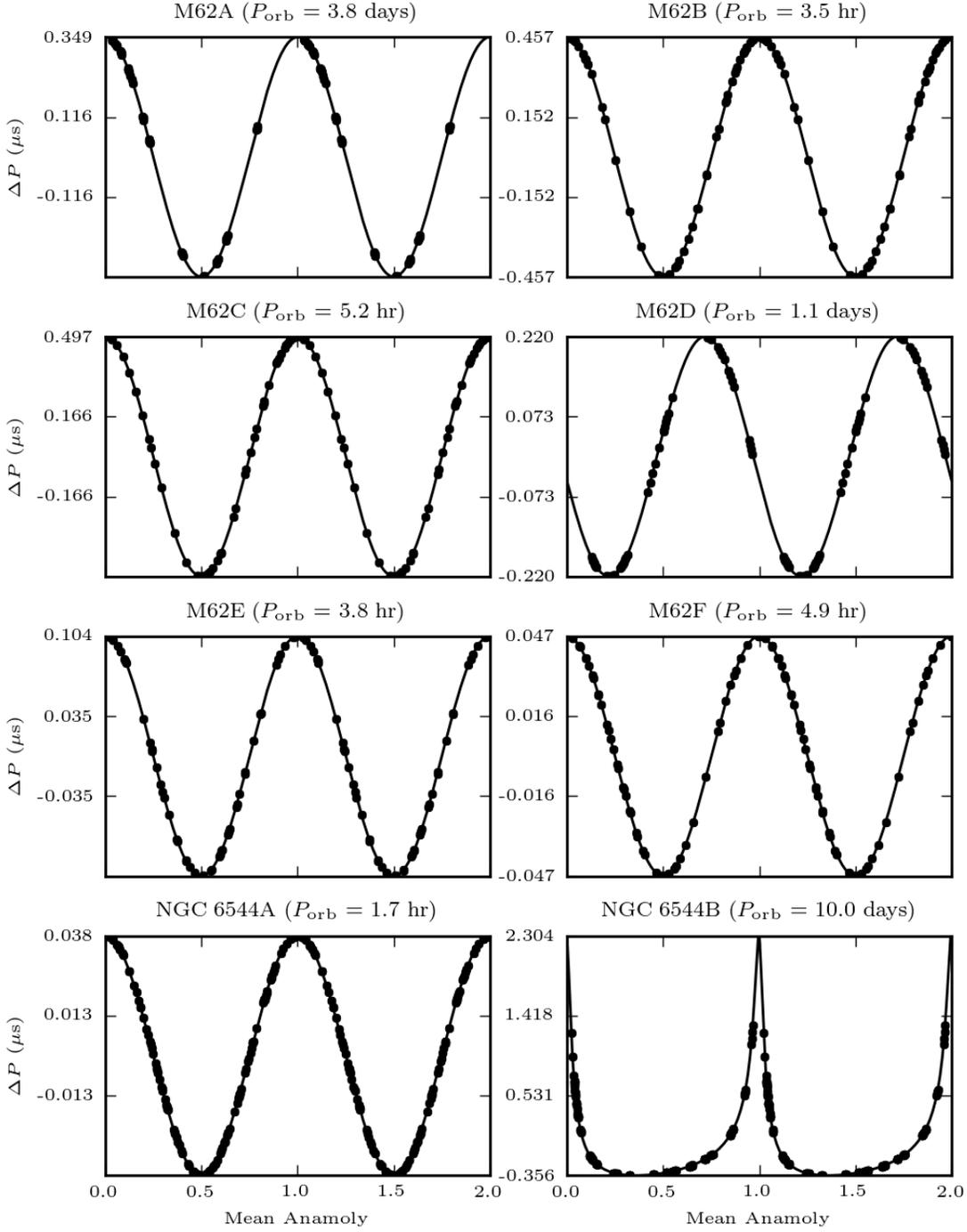}
\caption{The change in observed rotational period due to the Doppler
  effect as a function of mean anomaly.  The circles show the observed
  periods.  For clarity, the data are duplicated from orbital phase
  1--2.  Note that the high eccentricity of NGC~6544B causes the large
  deviation from a sinusoidal modulation of the period.
  \label{fig:dop_ps}}
\end{figure}

\begin{figure}
\centering
\includegraphics[width=5.75in]{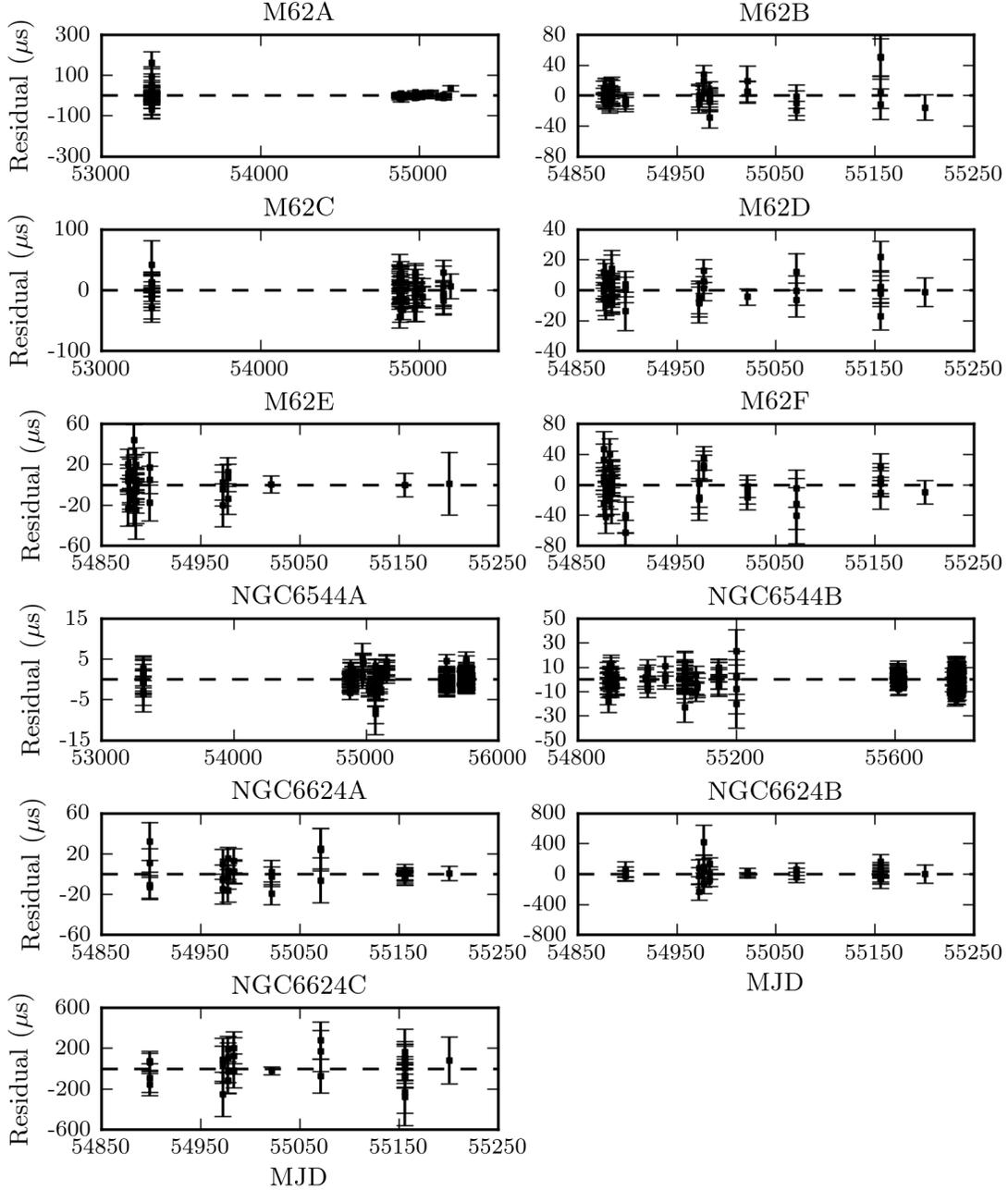}
\caption{Post-fit timing residuals for the pulsars in these clusters.
  Only unambiguously phase connected TOAs are displayed.  The TOAs
  around MJD 55600 for NGC~6544A and B are from recent Shapiro delay
  observations of NGC~6544B.  \label{fig:residuals}}
\end{figure}

All of our timing solutions can be found in Tables
\ref{table:M62_old}--\ref{table:NGC6624}, along with some derived
properties of the pulsars.  Average pulse profiles and Doppler
modulated pulse periods of the binary pulsars in these clusters are
shown in Figs.\ \ref{fig:profs} \& \ref{fig:dop_ps}.  Post-fit timing
residuals are shown in Fig. \ref{fig:residuals}.  We discuss each
individual system below.

\subsection{Pulsars with Previously Published Timing Solutions
  \label{sec:prev}}

PSRs J1701$-$3006A, B, and C and J1823$-$3021A and B all have timing
solutions published by other authors \citep{pdm+03,bbl+94}.  We
constructed our own timing solutions based solely on our data to
confirm that there were no irregularities in our data or methods.  We
find that nearly all our measured parameters agree with those
published elsewhere to within errors\footnote{Our measurement of right
  ascension for M62A differs from the results given in \citet{pdm+03}
  by $0.003"$, which is approximately five times the formal 1-$\sigma$
  errors}.  The one major exception to this is PSR J1701$-$3006B
(hereafter M62B), where we see a highly significant change in orbital
period and dispersion measure (DM) compared with the results of
\citet{pdm+03}.  We have also measured the rate of change of the
orbital period $\dot{P}\rmsub{b} = -5.51(62) \times 10^{-12}$.  We
performed an F-test to determine if the addition of $\dot{P}\rmsub{b}$
was indeed required by the data.  Without it, $\chi^2 = 142.82$ with
71 degrees of freedom, while after fitting for $\dot{P}\rmsub{b}$ the
$\chi^2$ improved to $73.94$ with 70 degrees of freedom.  The
probability that this improvement is due to chance is $< 1.3 \times
10^{-11}$, so the measured $\dot{P}\rmsub{b}$ does indeed seem to be
required by the data.  M62B is known to eclipse and has an optical and
X-ray counterpart \citep{cfp+08}.  As \citeauthor{cfp+08} discuss, the
pulsar and companion star are almost certainly interacting.  The
contribution to $\dot{P}\rmsub{b}$ expected from general relativity
(GR) is two orders of magnitude smaller than observed, assuming
$M\rmsub{p} = 1.4\; \Msun$ and $M\rmsub{c} = M\rmsub{c,min}$, and
there is no realistic combination of pulsar and companion masses and
inclination angles that would lead to such a large relativistic
$\dot{P}\rmsub{b}$.  As such, classic tidal effects of the extended
companion are probably the cause of the change in orbital period.  Our
RMS timing residuals are over a factor of two smaller than obtained by
\citet{pdm+03}, which probably explains why we were able to detect
$\dot{P}\rmsub{b}$ even though we had a shorter timing baseline.

\subsection{PSR J1701$-$3006D \label{sec:M62D}}

PSR J1701$-$3006D (hereafter M62D) is a $3.42\; \ms$ binary MSP.  M62D
(along with E and F) were discovered, and initial orbital solutions
were given, by \citet{cha03}.  The orbital period is $1.1\;
\mathrm{days}$, with a small eccentricity\footnote{We have used the
  ELL1 timing model in \texttt{TEMPO2}, which is appropriate when $a\,
  \sin{i}/c\, e^2 \ll 1$ \citep{lcw+01}.}, $e \sim 4.12 \times
10^{-4}$.  The minimum companion mass is $\sim 0.12\; \Msun$ (assuming
a $1.4\; \Msun$ MSP), and is likely a white dwarf.  We see no evidence
for eclipses in M62D, but none of our observations cover conjunction,
when an eclipse would be most likely.  Three observations do, however,
start or end within four hours ($\sim 15\%$ of the orbital period) of
conjunction, and one observation ends only 15 minutes before
conjunction.  The lack of eclipses increase our confidence that the
companion is a white dwarf, and not a main sequence (MS) star (for
which eclipses should be common).  We are unable to measure any
precession in the longitude of periastron, so no further constraints
can be placed on the mass or geometry of the system at this time.  The
acceleration of M62D ($\dot{P} P^{-1}$) is somewhat higher than for
the majority of GC pulsars \citep{ran08}, though not exceedingly so.
We note, though, that PSRs J1701$-$3006E and F have similarly large
accelerations.

\subsection{PSR J1701$-$3006E \label{sec:M62E}}

PSR J1701$-$3006E (hereafter M62E) is a $3.23\; \ms$ binary MSP in a
circular orbit.  The orbital period is only $3.80\; \hr$, and the
projected semi-major axis is only $0.0302\; \Rsun$.  When combined
with the low minimum companion mass ($0.031\; \Msun$, or only 31
Jupiter masses) and the presence of eclipses, M62E is clearly a black
widow pulsar \citep{kbr+05}.  The presence of eclipses makes it likely
that the system is being seen at a high inclination, so that the true
companion mass is probably close to the minimum.  We have good orbital
coverage of the system, and observe both ingress and egress during
eclipses, which are sharp and do not lead to a significant change in
\dm.  The eclipses are centered around orbital conjunction and occur
for $\sim 12\%$ of the orbit (or roughly 12 minutes).  We see no
evidence for eclipses at other orbital phases.  The projected extent
of the eclipsing material is $\sim 7 \times 10^5\; \km$.  A
relationship for the radius of the companion is given by
\citet{kin88}:
\begin{eqnarray}
R\rmsub{c} \simeq 10^4\; \km\: (1 + X)^{5/3} 
                  \left (\frac{M\rmsub{c}}{\Msun} \right)^{-1/3}
\end{eqnarray}
where $X$ is the hydrogen fraction.  For $X = 0.7$ we find $R\rmsub{c}
\sim 8 \times 10^4\; \km$, which is substantially smaller than the
implied size of the eclipsing region.  It seems likely that the
eclipses are being caused by an extended region of gas surrounding the
companion.

\subsection{PSR J1701$-$3006F \label{sec:M62F}}

PSR J1701$-$3006F (hereafter M62F) is a $2.29\; \ms$ binary MSP in a
circular orbit.  Like M62E, it has a low minimum mass companion ($\sim
22$ Jupiter masses) and occupies a region in
$M\rmsub{c,min}$-$P\rmsub{b}$ phase space typical of non-eclipsing
black widow pulsars.  Indeed, despite full orbital coverage we see no
evidence for flux variability as a function of orbital phase.
Nonetheless, given the very low mass limit and highly circularized
orbit, we still favor classifying M62F as a black widow, and the lack
of eclipses probably indicate that the system is not being viewed
close to edge-on \citep{fre05}.

\subsection{PSR J1807$-$2459A \label{sec:NGC6544A}}

PSR J1807$-$2459A (NGC~6544A) has already been discussed extensively
by \citet{dlm+01} and \citet{rgh+01}.  It is a $3.06\; \ms$ MSP in a
black widow system with a highly circular orbit and $M\rmsub{c,min} =
0.009\; \Msun$.  We were able to reliably phase connect the data
collected in 2009--2010 to data taken in 2011 as part of our Shapiro
delay observations of PSR J1807$-$2500B (see below), as well as to
older GBT data from 2004\footnote{The difference between the pre-fit
  residuals of the 2004 data and the predictions of our 2009--2011
  based timing solution are only 6\% of pulse phase.  Therefore, we
  are confident that we can phase connect all our data.}.  The orbital
parameters of this system were already well determined by previous
authors, however no phase-coherent timing solution had ever been
derived or published.  Our new timing solution includes, for the first
time, a precise measurement of the position of the system and its
spin-down.  The orbit in our solution is in good agreement with
previous analyses, but significantly more precise; it allowed us to
measured $\dot{P}\rmsub{b}$.  Like M62B, the GR contribution to
$\dot{P}\rmsub{b}$ is two orders of magnitude smaller than observed
for a $1.4\; \Msun$ pulsar and minimum mass companion.  However, there
are no eclipses observed for NGC~6544A, so it is possible that the
orbital inclination is significantly smaller than $90^{\circ}$, which
would imply a higher companion mass and larger GR contribution.
Nonetheless, for GR to contribute even 10\% of the measured
$\dot{P}\rmsub{b}$, the companion would have to have a mass of $\sim
0.06\; \Msun$, corresponding to $i = 9^{\circ}$.  For a distribution
of inclination angles that is flat in $\cos{i}$, there is only a $\sim
1\%$ probability of having $i < 9^{\circ}$.  It thus seems that we can
safely rule out a significant contribution from GR to
$\dot{P}\rmsub{b}$.  We can also rule out significant contamination
from acceleration in the cluster because the observed
$\dot{P}\rmsub{b}/P\rmsub{b}$ is several times larger than the maximum
cluster acceleration.  It thus seems likely that $\dot{P}\rmsub{b}$
can be explained by normal long-term black widow behavior
\citep{nat00}.  NGC~6544A is offset from the center of the cluster by
$4.6''$, or roughly $1.5$ core radii.  At the distance of NGC~6544,
this is only $0.06\; \pc$.

The observed $\dot{P} < 0$, which, if it were intrinsic to the pulsar,
would imply that the pulsar is spinning \emph{up}.  In reality,
$\dot{P}$ is contaminated by the acceleration of the pulsar within the
gravitational field of the cluster and Galaxy, and the fact that
$\dot{P} <0$ gives unambiguous evidence that the pulsar is on the far
side of the cluster and accelerating towards Earth.  This provides us
with a probe of the mass enclosed at the projected position of the
pulsar and the mass-to-light ratio ($\Upsilon$) \citep{phi93}.
Following \citet{dpf+02},
\begin{eqnarray}
  \Upsilon\rmsub{V} \geq 1.96 \times 10^{17}\: \frac{a\rmsub{c}}{c} 
  \left (\frac{\Sigma\rmsub{V}(<\theta_{\perp})}
    {10^4\; \mathrm{L_{\Sun,V}}\: \pc^{-2}} \right )^{-1}
\end{eqnarray}
where $\Sigma\rmsub{V}(<\theta_{\perp})$ is the mean surface
brightness interior the position of the pulsar.  To calculate
$\Sigma\rmsub{V}(<\theta_{\perp})$, we assume a constant surface
brightness in the core of the cluster.  The cluster acceleration,
$a\rmsub{c}$, is obtained by re-arranging Eq. \ref{eqn:pdot}.  In this
case, the intrinsic spin-down of the pulsar is estimated by using the
formula for characteristic age ($\tau\rmsub{c}$),
\begin{eqnarray}
  \frac{\dot{P}\rmsub{int}}{P} \approx \frac{1}{2 \tau\rmsub{c}},
\end{eqnarray}
assuming a pulsar age of $10\; \Gyr$.  We find $\dot{P}\rmsub{int}/P =
1.6\times 10^{-18}\; \ps$.  We were unable to find a proper motion for
NGC~6544 in the literature, but our results are not very sensitive to
this---for example, if the cluster had a transverse velocity of $100\;
\km\: \ps$, it would affect our results at the 10\% level.  Combining
all of this information, we find $\Upsilon\rmsub{V} \geq 0.072\;
\Msun/\Lsun$.  Massive stellar remnants in the cores of GCs should
give rise to a higher $\Upsilon\rmsub{V}$ (typically $\sim 3$--$4\;
\Msun/\Lsun$), so this result is very unconstraining.

\subsection{PSR J1807$-$2500B \label{sec:NGC6544B}}

PSR J1807$-$2500B \citep[NGC~6544B;][]{cha03} is the most intriguing
system in our sample.  As with NGC~6544A, were able to phase connect
data spanning 2004--2011\footnote{The difference between the pre-fit
  residuals of the 2004 data and the predictions of our 2009--2011
  based timing solution are $< 3\%$ of pulse phase.  Therefore, we are
  confident that we can phase connect all our data.}.  The pulsar has
a period of $4.19\; \ms$ and is in a highly eccentric orbit, with $e =
0.747$ (see Fig.  \ref{fig:dop_ps}).  The very high eccentricity of
this system has allowed us to measure a very significant precession of
periastron, $\dot{\omega} = 0.018319(12)^{\circ}\; \pyr$.  We have
also measured Shapiro delay in NGC~6544B\footnote{To the best of our
  knowledge, this is the first detection of Shapiro delay in a GC
  pulsar.}. We discuss both measurements in more detail below.

\subsubsection{Possible Contributions to $\mathbf{\dot{\omega}}$
  \label{sec:omdot}} 

The observed orbital precession could be due to any combination of
tidal deformation of the companion, spin-orbit coupling, or GR
effects, but for various reasons discussed below, we believe that it
is due almost entirely to GR.

Tidal deformation of the companion would require a MS or giant
companion.  This already seems unlikely based solely on the mass
function---the minimum companion mass for $M\rmsub{p} = 1.4\; \Msun$
is $\sim 1.2\; \Msun$, which is well above the turn-off mass in GCs
($\sim 0.8\; \Msun$).  Even if we use a smaller pulsar mass (say
$1.2\; \Msun$), the minimum companion mass is still $\sim 1.1\;
\Msun$.  This rules out a MS companion, unless it was a blue
straggler, but given how rare these are, this seems exceedingly
unlikely.  Any giants would have to be close to the turn-off mass and
would experience Roche lobe overflow.  The resulting gas in the system
would give rise to eclipses, especially if the orbit is highly
inclined (and as we show below, it is) and near conjunction.  As it
turns out, we see no evidence for eclipses at any orbital phase,
including conjunction, so a giant seems unlikely.  A blue straggler
would also probably cause eclipses or some other timing irregularities
in the pulsar.  In fact, NGC~6544B times remarkably well, with no
unmodeled trends in the data and no need for an error factor to bring
$\chi^2\rmsub{red} = 1$.  Hence, we can safely rule out a significant
contribution to $\dot{\omega}$ by tidal deformation.

Spin orbit coupling is a possibility if the companion is rapidly
rotating, but it scales as $|\dot{x}/x|$ times a geometrical factor,
where $x$ is the projected semi-major axis; the geometrical factor is
expected to be $<10$ in 80\% of cases \citep[e.g.][and references
therein]{frb+08}.  Our current best limit on $\dot{x}$ implies that
precession due to spin-orbit coupling should be smaller than our
extremely small measurement uncertainties in $\dot{\omega}$

For the reasons outlined above, we feel confident that the observed
$\dot{\omega}$ is due almost entirely to GR.  In this case the total
mass of the system is given by
\begin{eqnarray}
\frac{M\rmsub{tot}}{\Msun} = \left (\frac{\dot{\omega}}{3} \right)^{3/2}
                             \frac{(1 - e^2)^{3/2}}{T_{\Sun}}
                             \left (\frac{P\rmsub{b}}{2\pi} \right )^{5/2},
\end{eqnarray}
where $T_{\Sun} = G \Msun c^{-3}$, and $P\rmsub{b}$ is the orbital
period.  Using our measured values, we find $M\rmsub{tot} =
2.56763(59)\; \Msun$.

\subsubsection{Shapiro Delay Measurement \label{sec:shapiro}} 

The total system mass of NGC~6544B is fairly high for a pulsar binary
system.  When combined with the observed mass function, it implies a
high orbital inclination and massive companion, making NGC~6544B an
excellent candidate for a measurement of Shapiro delay.  We observed
the pulsar on ten occasions over a variety of orbital phases,
including two eight hour tracks at or near orbital
conjunction\footnote{Because the orbital period is almost exactly ten
  days, conjunction moves slowly in sidereal time, and as such is not
  visible from a given telescope for long stretches.  We were not able
  to schedule all our observations contiguously around conjunction due
  to this constraint.}.

We used the DDH timing model developed by \citet{fw10}, which
parameterizes the Shapiro delay as
\begin{eqnarray}
\varsigma & \equiv & \frac{s}{1 + \sqrt{1 - s^2}} \\
h_3       & \equiv & r \varsigma^3
\end{eqnarray}
where $s = \sin{i}$ and $r = \mathrm{T_{\Sun}} M\rmsub{c}$ are the
traditional Shapiro delay parameters.  In this parameterization, $h_3$
quantifies the amplitude of the third harmonic and $\varsigma$ the
ratio of successive harmonics of the Shapiro delay.  These parameters
are less correlated with each other and with other orbital parameters
than are $s$ and $r$, and provide a better description of the
combinations of orbital inclination and companion mass allowed by the
timing.  From the measured values of $h_3$ and $\varsigma$ we derive
$\sin{i} = 0.99715(20)$ and $M\rmsub{c} = 1.02(17)\; \Msun$.  On its
own, the Shapiro delay does not provide precise measurements of the
component masses, but the DDH model does not assume that
$\dot{\omega}$ is relativistic, and thus does not make full use of the
available information.

\begin{figure}
\centering
\includegraphics[width=5.75in]{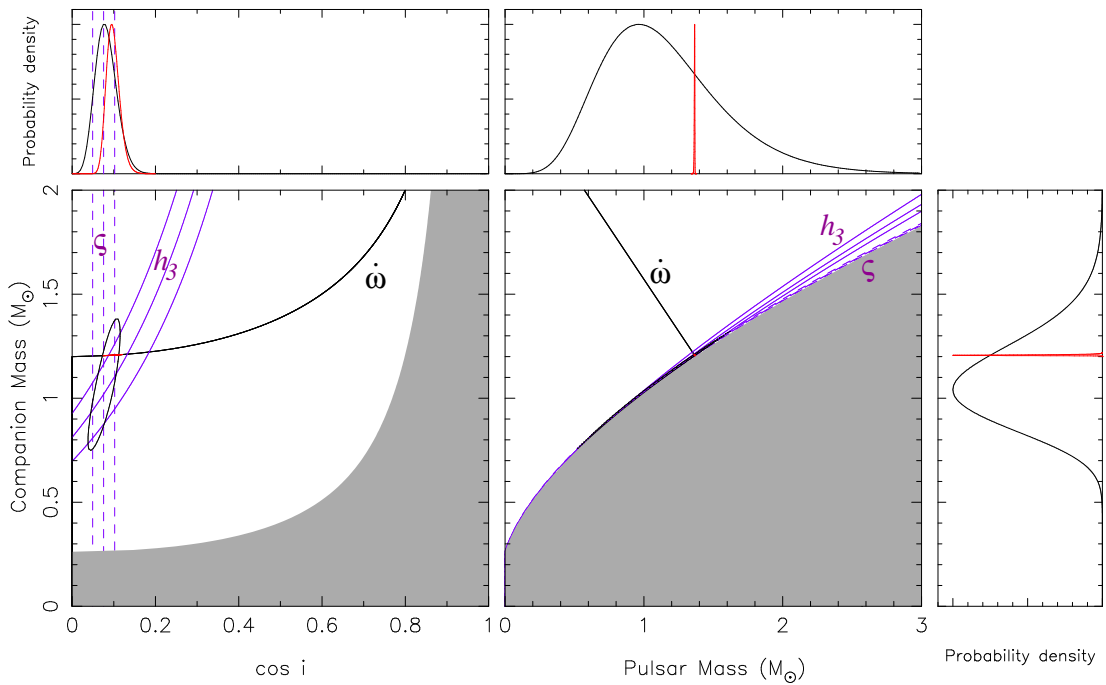}
\caption{Constraints on the inclination angle and masses in NGC~6544B
  from $\chi^2$ maps and the DDH timing model.  Our measurement of
  $\dot{\omega}$ implies a total mass of $2.57190(73)\; \Msun$,
  indicated by the black lines. The 1-$\sigma$ uncertainty is so small
  that it is invisible at this scale.  The solid purple lines indicate
  the constraints from $h_3$ and the dashed purple lines indicate the
  constraints from $\varsigma$. {\em Left}: $\cos i$-$M\rmsub{c}$
  plot.  The gray region is excluded by the condition $M\rmsub{p} >
  0$.  {\em Right}: $M\rmsub{p}$-$M\rmsub{c}$ plot.  The gray region
  is excluded by the condition $|\sin i| \leq 1$.  The black contours
  include 68.3\% of the total probability of a 2-D probability
  distribution function (PDF), calculated from a $\chi^2$ map that
  used only the Shapiro delay to constrain the masses. The red contour
  levels further assume that GR is correct and is the sole cause of
  $\dot{\omega}$.  {\em Top and right marginal plots}: 1-D PDFs for
  $\cos i$, $M\rmsub{p}$ and $M\rmsub{c}$, obtained by marginalization
  of both 2-D PDFs.  The second PDF (in red) yields median masses of
  $M\rmsub{p}^{\mathrm{med}} = 1.3649^{+0.0017}_{-0.0022}\; \Msun$,
  and $M\rmsub{c}^{\mathrm{med}} = 1.2068^{+0.0022}_{-0.0016}\;
  \Msun$.  The probability peaks at $M\rmsub{p}^{\mathrm{max}} =
  1.3654\; \Msun$ and $M\rmsub{c}^{\mathrm{max}} = 1.2063\; \Msun$.
  These are in excellent agreement with the masses obtained from the
  DDGR timing solution, given in Table \ref{table:NGC6544B}.
  \label{fig:NGC6544B_MvM}}
\end{figure}

To improve our mass estimates we used the DDGR model \citep{tw89},
which assumes that GR correctly describes the system.  In this model,
$M\rmsub{tot}$ and $M\rmsub{c}$ are free parameters and all
post-Keplerian (PK) parameters are derived from these measurements.
From this we obtain $M\rmsub{tot}^{\mathrm{DDGR}} = 2.57190(73)\;
\Msun$, $M\rmsub{c}^{\mathrm{DDGR}} = 1.2064(20)\; \Msun$, and derive
$M\rmsub{p}^{\mathrm{DDGR}} = 1.3655(21)\; \Msun$ (see Table
\ref{table:NGC6544B} for the complete solution).  This is the most
precise mass ever derived for an MSP, the previous being J1903+0327
\citep{fbw+11}.  To understand this exceptional precision and verify
it, we created $\chi^2$ maps from the DDH model, using the Keplerian
parameters of the DDGR solution as a starting point. For each
$\cos{i}$ and $M\rmsub{c}$, we calculated all five PK parameters
($\dot{\omega}$, $\gamma$, $\dot{P}\rmsub{b}$, $\varsigma$ and $h_3$)
assuming they are determined solely by GR; these were kept fixed while
\texttt{TEMPO2} fit for all other parameters.  The resulting $\chi^2$
values were recoded and the associated 2-D probability distribution
function (PDF) was calculated using the Bayesian analysis technique
discussed in detail in \citet{sna+02}.  This 2-D PDF was then
translated into the $M\rmsub{c}$-$M\rmsub{p}$ plane using the mass
function of our starting DDH solution.  The original 2-D PDF was then
collapsed onto the $\cos{i}$ and $M\rmsub{c}$ axes, with the derived
2-D PDF collapsed into the $M\rmsub{p}$ axis (see red lines in Fig.
\ref{fig:NGC6544B_MvM}), generating 1-D PDFs for the latter
quantities. These are slightly asymmetric; they have medians and $\pm
1 \sigma$ percentiles at: $\cos{i}^{\mathrm{med}} =
0.097^{+0.017}_{-0.015}$, $M\rmsub{p}^{\mathrm{med}} =
1.3649^{+0.0017}_{-0.0022}\; \Msun$, and $M\rmsub{c}^{\mathrm{med}} =
1.2068^{+0.0022}_{-0.0016}\; \Msun$.  The maximum probability occurs
at $\cos{i}^{\mathrm{max}} = 0.095$, $M\rmsub{p}^{\mathrm{max}} =
1.3654\; \Msun$, and $M\rmsub{c}^{\mathrm{max}} = 1.2063\; \Msun$.
These are in excellent agreement with the best-fit results of the DDGR
model, and confirm the value and the small uncertainty of the masses.

The regions of the $\cos{i}$-$M\rmsub{p}$ plane (and derived
$M\rmsub{p}$-$M\rmsub{c}$ plane) allowed by each measured PK parameter
are depicted in Fig. \ref{fig:NGC6544B_MvM}.  This allows us to
understand the reason for such a precise mass: the constraint provided
by $\dot{\omega}$ (the total mass $M\rmsub{tot}$, according to GR)
cuts the $\cos{i}$-$M\rmsub{c}$ (or $M\rmsub{p}$-$M\rmsub{c}$) regions
allowed by the Shapiro delay very sharply.

The measured companion mass strengthens the case made in
\S\ref{sec:omdot} that the companion must be a massive white dwarf or
second neutron star.  The companion is the most massive known around a
fully recycled pulsar.  If this is a double neutron star system, it is
only the second known in GCs, the first being M15C
\citep{agk+90,jcj+06}.  It is likely that the pulsar was recycled by a
different companion, which was then ejected in an exchange
interaction.  Deep Hubble Space Telescope images of the cluster may be
able to detect a white dwarf companion, while a non-detection of the
companion would strengthen the case for a double neutron star system.

We have measured three PK parameters ($\dot{\omega}$, $\varsigma$, and
$h_3$).  Assuming GR is correct, all three agree on the same region of
the $M\rmsub{c}$-$M\rmsub{p}$ plane.  As such, GR passes this test,
albeit at relatively low precision.  Our current best measurement of a
fourth PK parameter, the gravitational redshift, is $\gamma =
0.026(14)\; \s$.  This is consistent with the GR prediction of
$0.014\; \s$.  We plan to continue monitoring this system long-term,
and in a few years expect to have a more precise measurement of
$\gamma$ that will allow for a second test of GR.

\subsection{PSR J1823$-$3021C \label{sec:NGC6624C}}

PSR J1823$-$3021C (NGC~6624C) has $P = 0.406\; \s$, unusually slow
among globular cluster pulsars but, surprisingly, the second
such slow pulsar in NGC~6624. It was discovered by \citet{cha03}; we
have now measured, for the first time, $\dot{P} = 2.25 \times
10^{-16}\; \s\: \ps$.  Using Eq. \ref{eqn:amax}, we find
$\dot{P}\rmsub{c,max} = 1.7 \times 10^{-17}\; \s\: \ps$, over an order
of magnitude smaller than the measured $\dot{P}$.  The contributions
from the Galaxy and the Shklovskii effect are even smaller still, so
it is clear that the observed $\dot{P}$ is due almost entirely to the
intrinsic pulsar spin-down.  The implied characteristic age and
surface magnetic field is $\tau\rmsub{c} \sim 2.8\times 10^7\; \yr$
and $3.1 \times 10^{11}\; \gauss$.  This makes NGC~6624C, like
NGC~6624B, similar to the ``normal'', non-recycled pulsars (NRPs)
usually seen in the Galactic disk.  As explained in \S\ref{sec:intro},
NRPs are typically assumed to form in core collapse supernova, which
require massive stars that have not existed in GCs for billions of
years.  Since NRPs have lifetimes $\ll 10^{9}\; \yr$, core collapse
supernovae cannot explain the presence of NGC~6624C and pulsars like
it.  The leading alternative explanation is electron capture
supernovae (ECS).  The kick velocities ($v\rmsub{kick}$) that pulsars
receive when they form via ECS are not well known, though there is
evidence that $v\rmsub{kick} \sim 10\; \km\: \ps$
\citep{prp+02,kjh06,dbo+06,mtp09,wwk10}.  The escape velocity from the
center of NGC~6624 is $\sim 35\; \km\: \s$, which effectively places
an upper bound on the $v\rmsub{kick}$ that NGC~6624C received.  This
supports the notion that ECS kicks are much smaller than those induced
by core collapse supernovae.

GC NRPs can also enable useful statistical constraints on the
properties of ECS \citep{blt+11,ldr+11}.  NGC~6624C is one of only
four GC NRPs known, making it an important addition to this rare
family of pulsars.

\subsection{Pulsars Without Full Timing Solutions
  \label{sec:no_solutions}}

As part of this work, we discovered three previously unknown MSPs,
J1823-3021D, E, and F (hereafter NGC~6624D, E, and F).  However, we
were unable to obtain full timing solutions for them.  This was due to
the very low signal-to-noise ratio of the detections of all three
pulsars, due in part to their low flux densities, but also due to a
data processing error on our part.  As is typical, we de-dispersed the
raw data and combined many frequency channels to reduce data volume
(i.e. sub-banding), discarding the raw data afterward.  However, we
accidentally de-dispersed at the wrong DM for many of our
observations, thereby adding significant dispersive smearing and
degrading the signal from these already weak pulsars even further.
Without the raw data, we are unable to remedy this error.  PSR
J1823$-$3021A was bright enough that it could still be detected, and
the dispersive smearing was not large enough to significantly impact
the two long-period pulsars.  Despite this, we could still determine
some of the basic properties of NGC~6624D, E, and F.

We are able to confirm that NGC~6624D and E are isolated MSPs with $P
= 3.02$, and $4.39\; \ms$, respectively.  NGC~6624D was detected on
many occasions, but the TOA errors were quite large.  We were able to
measure the position with some accuracy, although without a reliable
measurement of $\dot{P}$, we are unable to break all the covariances
in the fit, so these coordinates should be used cautiously.  NGC~6624E
was only detected in three observations but there was no variation in
the apparent period of the pulsar between these observations, nor was
there any evidence of acceleration.  Both of these effects are
expected if the pulsar were in a binary system.  We were unable to
phase connect any of our observations due to large relative errors on
the measured period of the pulsar, and we cannot constrain its
position or $\dot{P}$.

NGC~6624F is an eclipsing binary MSP with $P = 4.85\; \ms$.  Like
NGC~6624E, NGC~6624F was only detected three times.  In each case, we
were able to clearly measure variations in $P$, $\dot{P}$ and
$\ddot{P}$ arising from orbital motion.  As it turns out, the apparent
period of the pulsar was similar and the acceleration was positive in
all three observations, indicating that the observations occurred at
overlapping orbital phases near inferior conjunction.  Because the
pulsar had the \emph{exact} same apparent period at some point during
each of our observations, there must have been an integer number of
orbits between these times.  We calculated the exact moment during
each observation when the pulsar had a reference barycentric period of
precisely $4.85\; \ms$ through a Taylor expansion of $P$, $\dot{P}$,
and $\ddot{P}$ and a bisection method.  An orbital period of $0.8827$
days implies exactly $106.0003$ and $146.9996$ complete orbits between
our three observations.  The fact that we were able to fit such a
precise integer number of orbits gives us confidence that this is the
true orbital period of the system.  If NGC~6624F is in a circular
orbit, each observation would mark points on an ellipse in the
$P$-$\dot{P}$ plane \citep{fkl01}.  With three observations and a good
determination of the spin and orbital period, we can, in principal,
uniquely constrain the equation of this ellipse, which in turn can
provide the intrinsic $P$, $P\rmsub{b}$, and $a \sin{i}/c$.  We
attempted to perform this fit to our data using a least squares
minimization routine and $0.01\; \mathrm{lt-s} \leq a \sin{i}/c \leq
1000\; \mathrm{lt-s}$.  We found a best-fit $a \sin{i}/c \approx 4.4\;
\mathrm{lt-s}$, but this solution has poor predictive power (i.e.,
when we folded our data using this solution, the pulsar was not
detected).  We were also unable to construct a coherent timing
solution in \texttt{TEMPO2} using this starting orbital solution.  We
believe that the lack of coverage at multiple orbital phases is
limiting our ability to accurately determine $a \sin{i}/c$.  It is
also possible that NGC~6624F is not in a circular orbit.

\section{Conclusion \label{sec:conc}}

We set out to obtain full timing solutions for nine previously known
GC pulsars in M62, NGC~6544, and NGC~6624 (three of which have never
been published).  Our efforts were successful for six of these
pulsars, and a seventh now has a partial solution.  We have confirmed
the nature of three black widow pulsars (M62E and F, and NGC~6544A).
NGC~6544A has $\dot{P} < 0$, providing unambiguous evidence that
$\dot{P}$ is dominated by acceleration in the cluster potential, and
providing us with a probe of $\Upsilon\rmsub{V}$ in the cluster core.
We find $\Upsilon\rmsub{V} \gtrsim 0.072$ which, while not very
constraining, is consistent with small amounts of low-luminosity
matter such as massive black holes or other stellar remnants.  We are
also able to confirm that NGC~6624C belongs to a rare class of
non-recycled GC pulsars similar to the slow pulsars found in the
Galactic disk.

The highlight of our work is unquestionably NGC~6544B.  This binary
MSP is in a highly eccentric orbit and exhibits clear orbital
precession and Shapiro delay.  Under the well justified assumption
that this is due almost entirely to GR, we are able to obtain a
most-probable total system mass of $2.57190(73)\; \Msun$, companion
mass of $1.2064(20)\; \Msun$, and pulsar mass of $1.3655(21)\; \Msun$.
This is the highest mass companion known to orbit a fully recycled
MSP, and raises the possibility that NGC~6544B is part of either
double neutron star system, which could form via exchange interactions
in the GC.

We are deeply grateful to Andrea Possenti and Alessandro Corongiu for
providing us with data collected by the Parkes Telescope that was used
to check for phase connection in our solutions for M62D, E, and F.  We
would also like to thank an anonymous referee for helpful comments.
R.\ Lynch acknowledges support from the GBT Student Support program
and the National Science Foundation grant AST-0907967 during the
course of this work.  The National Radio Astronomy Observatory is a
facility of the National Science Foundation operated under cooperative
agreement by Associated Universities, Inc.

\bibliographystyle{apj}
\bibliography{ms}{}

\begin{thebibliography}{51}
\expandafter\ifx\csname natexlab\endcsname\relax\def\natexlab#1{#1}\fi

\bibitem[{{Alpar} {et~al.}(1982){Alpar}, {Cheng}, {Ruderman}, \&
  {Shaham}}]{acr+82}
{Alpar}, M.~A., {Cheng}, A.~F., {Ruderman}, M.~A., \& {Shaham}, J. 1982, \nat,
  300, 728

\bibitem[{{Anderson} {et~al.}(1990){Anderson}, {Gorham}, {Kulkarni}, {Prince},
  \& {Wolszczan}}]{agk+90}
{Anderson}, S.~B., {Gorham}, P.~W., {Kulkarni}, S.~R., {Prince}, T.~A., \&
  {Wolszczan}, A. 1990, \nat, 346, 42

\bibitem[{{Biggs} {et~al.}(1994){Biggs}, {Bailes}, {Lyne}, {Goss}, \&
  {Fruchter}}]{bbl+94}
{Biggs}, J.~D., {Bailes}, M., {Lyne}, A.~G., {Goss}, W.~M., \& {Fruchter},
  A.~S. 1994, \mnras, 267, 125

\bibitem[{{Boyles} {et~al.}(2011){Boyles}, {Lorimer}, {Turk}, {Mnatsakanov},
  {Lynch}, {Ransom}, {Freire}, \& {Belczynski}}]{blt+11}
{Boyles}, J., {Lorimer}, D.~R., {Turk}, P.~J., {Mnatsakanov}, R., {Lynch},
  R.~S., {Ransom}, S.~M., {Freire}, P.~C., \& {Belczynski}, K. 2011, ArXiv
  e-prints

\bibitem[{{Camilo} \& {Rasio}(2005)}]{cr05}
{Camilo}, F. \& {Rasio}, F.~A. 2005, in Astronomical Society of the Pacific
  Conference Series, Vol. 328, Binary Radio Pulsars, ed. {F.~A.~Rasio \&
  I.~H.~Stairs}, 147

\bibitem[{{Carretta} {et~al.}(2000){Carretta}, {Gratton}, {Clementini}, \&
  {Fusi Pecci}}]{cgcf00}
{Carretta}, E., {Gratton}, R.~G., {Clementini}, G., \& {Fusi Pecci}, F. 2000,
  \apj, 533, 215

\bibitem[{{Chandler}(2003)}]{cha03}
{Chandler}, A.~M. 2003, PhD thesis, California Institute Of Technology

\bibitem[{{Cocozza} {et~al.}(2008){Cocozza}, {Ferraro}, {Possenti}, {Beccari},
  {Lanzoni}, {Ransom}, {Rood}, \& {D'Amico}}]{cfp+08}
{Cocozza}, G., {Ferraro}, F.~R., {Possenti}, A., {Beccari}, G., {Lanzoni}, B.,
  {Ransom}, S., {Rood}, R.~T., \& {D'Amico}, N. 2008, \apjl, 679, L105

\bibitem[{{D'Amico} {et~al.}(2001{\natexlab{a}}){D'Amico}, {Lyne},
  {Manchester}, {Possenti}, \& {Camilo}}]{dlm+01}
{D'Amico}, N., {Lyne}, A.~G., {Manchester}, R.~N., {Possenti}, A., \& {Camilo},
  F. 2001{\natexlab{a}}, \apjl, 548, L171

\bibitem[{{D'Amico} {et~al.}(2002){D'Amico}, {Possenti}, {Fici}, {Manchester},
  {Lyne}, {Camilo}, \& {Sarkissian}}]{dpf+02}
{D'Amico}, N., {Possenti}, A., {Fici}, L., {Manchester}, R.~N., {Lyne}, A.~G.,
  {Camilo}, F., \& {Sarkissian}, J. 2002, \apjl, 570, L89

\bibitem[{{D'Amico} {et~al.}(2001{\natexlab{b}}){D'Amico}, {Possenti},
  {Manchester}, {Sarkissian}, {Lyne}, \& {Camilo}}]{dpm+01}
{D'Amico}, N., {Possenti}, A., {Manchester}, R.~N., {Sarkissian}, J., {Lyne},
  A.~G., \& {Camilo}, F. 2001{\natexlab{b}}, \apjl, 561, L89

\bibitem[{{Dessart} {et~al.}(2006){Dessart}, {Burrows}, {Ott}, {Livne}, {Yoon},
  \& {Langer}}]{dbo+06}
{Dessart}, L., {Burrows}, A., {Ott}, C.~D., {Livne}, E., {Yoon}, S., \&
  {Langer}, N. 2006, \apj, 644, 1063

\bibitem[{{DuPlain} {et~al.}(2008){DuPlain}, {Ransom}, {Demorest}, {Brandt},
  {Ford}, \& {Shelton}}]{drd+08}
{DuPlain}, R., {Ransom}, S., {Demorest}, P., {Brandt}, P., {Ford}, J., \&
  {Shelton}, A.~L. 2008, in Society of Photo-Optical Instrumentation Engineers
  (SPIE) Conference Series, Vol. 7019, Society of Photo-Optical Instrumentation
  Engineers (SPIE) Conference Series, 70191D--70191D--10

\bibitem[{{Edwards} {et~al.}(2006){Edwards}, {Hobbs}, \& {Manchester}}]{ehm06}
{Edwards}, R.~T., {Hobbs}, G.~B., \& {Manchester}, R.~N. 2006, \mnras, 372,
  1549

\bibitem[{{Freire} {et~al.}(2001){Freire}, {Kramer}, \& {Lyne}}]{fkl01}
{Freire}, P.~C., {Kramer}, M., \& {Lyne}, A.~G. 2001, \mnras, 322, 885

\bibitem[{{Freire}(2005)}]{fre05}
{Freire}, P.~C.~C. 2005, in Astronomical Society of the Pacific Conference
  Series, Vol. 328, Binary Radio Pulsars, ed. {F.~A.~Rasio \& I.~H.~Stairs},
  405

\bibitem[{{Freire} {et~al.}(2011){Freire}, {Bassa}, {Wex}, {Stairs},
  {Champion}, {Ransom}, {Lazarus}, {Kaspi}, {Hessels}, {Kramer}, {Cordes},
  {Verbiest}, {Podsiadlowski}, {Nice}, {Deneva}, {Lorimer}, {Stappers},
  {McLaughlin}, \& {Camilo}}]{fbw+11}
{Freire}, P.~C.~C., {Bassa}, C.~G., {Wex}, N., {Stairs}, I.~H., {Champion},
  D.~J., {Ransom}, S.~M., {Lazarus}, P., {Kaspi}, V.~M., {Hessels}, J.~W.~T.,
  {Kramer}, M., {Cordes}, J.~M., {Verbiest}, J.~P.~W., {Podsiadlowski}, P.,
  {Nice}, D.~J., {Deneva}, J.~S., {Lorimer}, D.~R., {Stappers}, B.~W.,
  {McLaughlin}, M.~A., \& {Camilo}, F. 2011, \mnras, 412, 2763

\bibitem[{{Freire} {et~al.}(2005){Freire}, {Hessels}, {Nice}, {Ransom},
  {Lorimer}, \& {Stairs}}]{fhn+05}
{Freire}, P.~C.~C., {Hessels}, J.~W.~T., {Nice}, D.~J., {Ransom}, S.~M.,
  {Lorimer}, D.~R., \& {Stairs}, I.~H. 2005, \apj, 621, 959

\bibitem[{{Freire} {et~al.}(2008){Freire}, {Ransom}, {B{\'e}gin}, {Stairs},
  {Hessels}, {Frey}, \& {Camilo}}]{frb+08}
{Freire}, P.~C.~C., {Ransom}, S.~M., {B{\'e}gin}, S., {Stairs}, I.~H.,
  {Hessels}, J.~W.~T., {Frey}, L.~H., \& {Camilo}, F. 2008, \apj, 675, 670

\bibitem[{{Freire} {et~al.}(2007){Freire}, {Ransom}, \& {Gupta}}]{frg07}
{Freire}, P.~C.~C., {Ransom}, S.~M., \& {Gupta}, Y. 2007, \apj, 662, 1177

\bibitem[{{Freire} \& {Wex}(2010)}]{fw10}
{Freire}, P.~C.~C. \& {Wex}, N. 2010, \mnras, 409, 199

\bibitem[{{Harris}(1996)}]{har96}
{Harris}, W.~E. 1996, \aj, 112, 1487 (arXiv:1012.3224)

\bibitem[{{Haslam} {et~al.}(1982){Haslam}, {Salter}, {Stoffel}, \&
  {Wilson}}]{hss+82}
{Haslam}, C.~G.~T., {Salter}, C.~J., {Stoffel}, H., \& {Wilson}, W.~E. 1982,
  \aaps, 47, 1

\bibitem[{{Hessels} {et~al.}(2006){Hessels}, {Ransom}, {Stairs}, {Freire},
  {Kaspi}, \& {Camilo}}]{hrs+06}
{Hessels}, J.~W.~T., {Ransom}, S.~M., {Stairs}, I.~H., {Freire}, P.~C.~C.,
  {Kaspi}, V.~M., \& {Camilo}, F. 2006, Science, 311, 1901

\bibitem[{{Hotan} {et~al.}(2004){Hotan}, {van Straten}, \&
  {Manchester}}]{hvm04}
{Hotan}, A.~W., {van Straten}, W., \& {Manchester}, R.~N. 2004, \pasa, 21, 302

\bibitem[{{Jacoby} {et~al.}(2006){Jacoby}, {Cameron}, {Jenet}, {Anderson},
  {Murty}, \& {Kulkarni}}]{jcj+06}
{Jacoby}, B.~A., {Cameron}, P.~B., {Jenet}, F.~A., {Anderson}, S.~B., {Murty},
  R.~N., \& {Kulkarni}, S.~R. 2006, \apjl, 644, L113

\bibitem[{{Kaplan} {et~al.}(2005){Kaplan}, {Escoffier}, {Lacasse}, {O'Neil},
  {Ford}, {Ransom}, {Anderson}, {Cordes}, {Lazio}, \& {Kulkarni}}]{kel+05}
{Kaplan}, D.~L., {Escoffier}, R.~P., {Lacasse}, R.~J., {O'Neil}, K., {Ford},
  J.~M., {Ransom}, S.~M., {Anderson}, S.~B., {Cordes}, J.~M., {Lazio},
  T.~J.~W., \& {Kulkarni}, S.~R. 2005, \pasp, 117, 643

\bibitem[{{King}(1988)}]{kin88}
{King}, A.~R. 1988, \qjras, 29, 1

\bibitem[{{King} {et~al.}(2005){King}, {Beer}, {Rolfe}, {Schenker}, \&
  {Skipp}}]{kbr+05}
{King}, A.~R., {Beer}, M.~E., {Rolfe}, D.~J., {Schenker}, K., \& {Skipp}, J.~M.
  2005, \mnras, 358, 1501

\bibitem[{{Kitaura} {et~al.}(2006){Kitaura}, {Janka}, \& {Hillebrandt}}]{kjh06}
{Kitaura}, F.~S., {Janka}, H., \& {Hillebrandt}, W. 2006, \aap, 450, 345

\bibitem[{{Lange} {et~al.}(2001){Lange}, {Camilo}, {Wex}, {Kramer}, {Backer},
  {Lyne}, \& {Doroshenko}}]{lcw+01}
{Lange}, C., {Camilo}, F., {Wex}, N., {Kramer}, M., {Backer}, D.~C., {Lyne},
  A.~G., \& {Doroshenko}, O. 2001, \mnras, 326, 274

\bibitem[{{Lynch} {et~al.}(in prep.){Lynch}, {Lorimer}, {Ransom}, \&
  {Boyles}}]{ldr+11}
{Lynch}, R.~S., {Lorimer}, D.~D., {Ransom}, S.~M., \& {Boyles}, J. in prep.,
  \apj

\bibitem[{{Lyne} {et~al.}(1996){Lyne}, {Manchester}, \& {D'Amico}}]{lmd96}
{Lyne}, A.~G., {Manchester}, R.~N., \& {D'Amico}, N. 1996, \apjl, 460, L41

\bibitem[{{Martin} {et~al.}(2009){Martin}, {Tout}, \& {Pringle}}]{mtp09}
{Martin}, R.~G., {Tout}, C.~A., \& {Pringle}, J.~E. 2009, \mnras, 397, 1563

\bibitem[{{Nice} {et~al.}(2000){Nice}, {Arzoumanian}, \& {Thorsett}}]{nat00}
{Nice}, D.~J., {Arzoumanian}, Z., \& {Thorsett}, S.~E. 2000, in Astronomical
  Society of the Pacific Conference Series, Vol. 202, IAU Colloq. 177: Pulsar
  Astronomy - 2000 and Beyond, ed. {M.~Kramer, N.~Wex, \& R.~Wielebinski}, 67

\bibitem[{{Nomoto}(1984)}]{nom84}
{Nomoto}, K. 1984, \apj, 277, 791

\bibitem[{{Nomoto}(1987)}]{nom87}
---. 1987, \apj, 322, 206

\bibitem[{{Pfahl} {et~al.}(2002){Pfahl}, {Rappaport}, {Podsiadlowski}, \&
  {Spruit}}]{prp+02}
{Pfahl}, E., {Rappaport}, S., {Podsiadlowski}, P., \& {Spruit}, H. 2002, \apj,
  574, 364

\bibitem[{{Phinney}(1992)}]{phi92}
{Phinney}, E.~S. 1992, Royal Society of London Philosophical Transactions
  Series A, 341, 39

\bibitem[{{Phinney}(1993)}]{phi93}
{Phinney}, E.~S. 1993, in Astronomical Society of the Pacific Conference
  Series, Vol.~50, Structure and Dynamics of Globular Clusters, ed.
  {S.~G.~Djorgovski \& G.~Meylan}, 141

\bibitem[{{Possenti} {et~al.}(2003){Possenti}, {D'Amico}, {Manchester},
  {Camilo}, {Lyne}, {Sarkissian}, \& {Corongiu}}]{pdm+03}
{Possenti}, A., {D'Amico}, N., {Manchester}, R.~N., {Camilo}, F., {Lyne},
  A.~G., {Sarkissian}, J., \& {Corongiu}, A. 2003, \apj, 599, 475

\bibitem[{{Ransom}(2008)}]{ran08}
{Ransom}, S.~M. 2008, in American Institute of Physics Conference Series, Vol.
  983, 40 Years of Pulsars: Millisecond Pulsars, Magnetars and More, ed.
  {C.~Bassa, Z.~Wang, A.~Cumming, \& V.~M.~Kaspi}, 415--423

\bibitem[{{Ransom} {et~al.}(2001){Ransom}, {Greenhill}, {Herrnstein},
  {Manchester}, {Camilo}, {Eikenberry}, \& {Lyne}}]{rgh+01}
{Ransom}, S.~M., {Greenhill}, L.~J., {Herrnstein}, J.~R., {Manchester}, R.~N.,
  {Camilo}, F., {Eikenberry}, S.~S., \& {Lyne}, A.~G. 2001, \apjl, 546, L25

\bibitem[{{Ransom} {et~al.}(2005){Ransom}, {Hessels}, {Stairs}, {Freire},
  {Camilo}, {Kaspi}, \& {Kaplan}}]{rhs+05}
{Ransom}, S.~M., {Hessels}, J.~W.~T., {Stairs}, I.~H., {Freire}, P.~C.~C.,
  {Camilo}, F., {Kaspi}, V.~M., \& {Kaplan}, D.~L. 2005, Science, 307, 892

\bibitem[{{Shklovskii}(1970)}]{shk70}
{Shklovskii}, I.~S. 1970, \sovast, 13, 562

\bibitem[{{Splaver} {et~al.}(2002){Splaver}, {Nice}, {Arzoumanian}, {Camilo},
  {Lyne}, \& {Stairs}}]{sna+02}
{Splaver}, E.~M., {Nice}, D.~J., {Arzoumanian}, Z., {Camilo}, F., {Lyne},
  A.~G., \& {Stairs}, I.~H. 2002, \apj, 581, 509

\bibitem[{{Taylor} \& {Weisberg}(1989)}]{tw89}
{Taylor}, J.~H. \& {Weisberg}, J.~M. 1989, \apj, 345, 434

\bibitem[{{Valenti} {et~al.}(2011){Valenti}, {Origlia}, \& {Rich}}]{vor11}
{Valenti}, E., {Origlia}, L., \& {Rich}, R.~M. 2011, \mnras, 560

\bibitem[{{Verbiest} {et~al.}(2008){Verbiest}, {Bailes}, {van Straten},
  {Hobbs}, {Edwards}, {Manchester}, {Bhat}, {Sarkissian}, {Jacoby}, \&
  {Kulkarni}}]{vbv+08}
{Verbiest}, J.~P.~W., {Bailes}, M., {van Straten}, W., {Hobbs}, G.~B.,
  {Edwards}, R.~T., {Manchester}, R.~N., {Bhat}, N.~D.~R., {Sarkissian}, J.~M.,
  {Jacoby}, B.~A., \& {Kulkarni}, S.~R. 2008, \apj, 679, 675

\bibitem[{{Webbink}(1985)}]{web85}
{Webbink}, R.~F. 1985, in IAU Symposium, Vol. 113, Dynamics of Star Clusters,
  ed. {J.~Goodman \& P.~Hut}, 541--577

\bibitem[{{Wong} {et~al.}(2010){Wong}, {Willems}, \& {Kalogera}}]{wwk10}
{Wong}, T., {Willems}, B., \& {Kalogera}, V. 2010, \apj, 721, 1689

\end{thebibliography}

\begin{deluxetable}{lccc} 
 \centering
  \tabletypesize{\footnotesize}
  \tablewidth{0pt}
  \tablecolumns{4}
  \tablecaption{Summary of Pulsars in this Study \label{table:psrs}}
  \tablehead{
    \colhead{Pulsar}            &
    \colhead{Previous Full}     &
    \colhead{New Full}          &
    \colhead{Reference}        \\
    \colhead{}                  &
    \colhead{ Timing Solution?} &
    \colhead{ Timing Solution?} &
    \colhead{}}
  \startdata
  \cutinhead{M62}
  J1701$-$3006A & Yes & Yes & A,B         \\
  J1701$-$3006B & Yes & Yes & B           \\
  J1701$-$3006C & Yes & Yes & B           \\
  J1701$-$3006D & No  & Yes & C           \\
  J1701$-$3006E & No  & Yes & C           \\
  J1701$-$3006F & No  & Yes & C           \\
  \cutinhead{NGC~6544}
  J1807$-$2459A & No  & Yes & A,D         \\
  J1807$-$2500B & No  & Yes & C           \\
  \cutinhead{NGC~6624}
  J1823$-$3021A & Yes & Yes & E           \\
  J1823$-$3021B & Yes & Yes & E           \\
  J1823$-$3021C & No  & Yes & C           \\
  J1823$-$3021D & No  & Yes & Unpublished \\
  J1823$-$3021E & No  & No  & Unpublished \\
  J1823$-$3021F & No  & No  & Unpublished \\
  \enddata  
  \tablerefs{A---\citet{dlm+01}; B---\citet{pdm+03};
    C---\citet{cha03}; D---\citet{rgh+01}; E---\citet{bbl+94}}
\end{deluxetable}

\begin{deluxetable}{lcccccc}
  \centering
  \tabletypesize{\footnotesize}
  \tablewidth{0pt}
  \tablecolumns{7}
  \tablecaption{Globular Cluster Properties \label{table:gcs}}
  \tablehead{
    \colhead{ID}                              &
    \colhead{$\ell$\tablenotemark{a}}         &
    \colhead{$b$\tablenotemark{b}}            &
    \colhead{$D$\tablenotemark{c}}            &
    \colhead{$r\rmsub{c}$\tablenotemark{d}}   &
    \colhead{$\mu\rmsub{V}$\tablenotemark{e}} &
    \colhead{$\sigma_v$\tablenotemark{f}}    \\
    \colhead{}                                &
    \colhead{(deg)}                           &
    \colhead{(deg)}                           &
    \colhead{($\kpc$)}                        &
    \colhead{(arcmin)}                        &
    \colhead{($\mathrm{mag\; arcsec^{-2}}$)}  &
    \colhead{($\km\: \ps$)}}
  \startdata
  M62      & 353.57 & 7.32    & 6.7 & 0.22 & 15.10 & 13.3 \\
  NGC~6544 & 5.84   & $-2.20$ & 3.0 & 0.05 & 16.31 & 5.89 \\
  NGC~6624 & 2.78   & $-7.91$ & 7.8 & 0.06 & 15.32 & 6.0  \\
  \enddata
  \tablenotetext{a}{Galactic longitude}
  \tablenotetext{b}{Galactic latitude}
  \tablenotetext{c}{Distance}
  \tablenotetext{d}{Core radius}
  \tablenotetext{e}{Central V-band surface brightness}
  \tablenotetext{f}{Central velocity dispersion}
  \tablecomments{All properties are taken from \citet[2010
    edition]{har96}, except for $\sigma_v$ in M62 and NGC~6544
    \citep{web85}, and NGC~6624 \citep{vor11}.}
\end{deluxetable}

\begin{deluxetable}{lccc}
  \rotate
  \centering
  \tabletypesize{\scriptsize}
  \tablewidth{0pt}
  \tablecolumns{4}
  \tablecaption{Parameters of the Previously Timed Pulsars in M62 
    \label{table:M62_old}}
  \tablehead{
    \colhead{Parameter} &
    \colhead{J1701$-$3006A} &
    \colhead{J1701$-$3006B} &
    \colhead{J1701$-$3006C}}
  \startdata
    \cutinhead{Timing Parameters}
    Right Ascension (J2000)                       \dotfill & 17:01:12.50942(49)            & 17:01:12.6677(10)             & 17:01:12.86592(35)            \\
    Declination (J2000)                           \dotfill & $-$30:06:30.173(36)           & $-$30:06:49.044(76)           & $-$30:06:59.415(21)           \\
    Spin Period ($\s$)                            \dotfill & 0.0052415662037958(29)        & 0.0035938521270515(34)        & 0.0076128487111379(68)        \\
    Period Derivative ($\s\: \ps$)                \dotfill & $-1.301(15) \times 10^{-19}$  & $-3.483(20) \times 10^{-19}$  & $-6.413(10) \times 10^{-20}$  \\
    Dispersion Measure ($\dmu$)                   \dotfill & 114.9654(84)                  & 115.21(38)                    & 114.5619(69)                  \\
    Reference Epoch (MJD)                         \dotfill & 55038.0                       & 55038.0                       & 55038.0                       \\
    Span of Timing Data (MJD)                     \dotfill & 53150--55200                  & 53315--55200                  & 53315--55200                  \\
    Number of TOAs                                \dotfill & 80                            & 81                            & 74                            \\
    RMS Timing Residual ($\us$)                   \dotfill & 6.4                           & 11.3                          & 15.1                          \\
    Error Factor                                  \dotfill & 1.6                           & 1.5                           & 1.2                           \\
    \cutinhead{Binary Parameters}
    Orbital Period (days)                         \dotfill & 3.8059483732(58)              & 0.14454541718(58)             & 0.21500007119(51)             \\
    Projected Semi-major Axis (lt-s)              \dotfill & 3.4837397(19)                 & 0.2527565(28)                 & 0.1928816(33)                 \\
    Epoch of Periastron Passage (MJD)             \dotfill & 55040.03822060(36)            & 55038.62576448(54)            & 55037.92664751(74)            \\
    Orbital Eccentricity                          \dotfill & $< 1.8 \times 10^{-6}$        & $< 4.5 \times 10^{-5}$        & $< 7.9 \times 10^{-5}$        \\
    Longitude of Periastron Passage (deg)         \dotfill & \nodata                       & \nodata                       & \nodata                       \\
    Rate of Change of Orbital Period ($10^{-12}$) \dotfill & \nodata                       & -5.12(62)$\times$10$^{-12}$   & \nodata                       \\
    \cutinhead{Derived Parameters}
    Mass Function $(\Msun)$                       \dotfill & 0.0031339643(51)              & 0.000829814(27)               & 0.0001666778(87)              \\
    Minimum Companion Mass ($\Msun$)              \dotfill & 0.2                           & 0.12                          & 0.071                         \\
    Offset from Cluster Center (core radii)       \dotfill & 1.50                          & 0.138                         & 0.773                         \\
    Intrinsic Spin-down ($\s\: \ps$)              \dotfill & $\leq 3.13 \times 10^{-19}$   & $\leq 4.4 \times 10^{-20}$    & $\leq 3.78 \times 10^{-19}$   \\
    Surface Magnetic Field ($10^9$ Gauss)         \dotfill & $\leq 1.3$                    & $\leq 0.4$                    & $\leq 1.7$                    \\
    Spin-down Luminosity ($10^{34}\; \erg\: \ps$) \dotfill & $\leq 8.6$                    & $\leq 3.7$                    & $\leq 3.4$                    \\
    Characteristic Age ($10^9\; \yr)$             \dotfill & $\geq 0.27$                   & $\geq 1.3$                    & $\geq 0.32$                   \\
    Mean $2\; \GHz$ Flux Density ($\uJy$)         \dotfill & 117                           & 78                            & 72                            \\
  \enddata
  \tablecomments{Numbers in parentheses represent 1-$\sigma$
  uncertainties in the last digits as determined by \texttt{TEMPO2},
  scaled such that the $\chi^2\rmsub{red} = 1$.  All timing solutions
  use the DE405 Solar System Ephemeris and the TT(BIPM2011) time standard.
  All values are reported in Barycentric Dynamical Time (TDB) units.
  Mean flux density estimates have a typical 10\%--20\% relative
  error.  Minimum companion masses were calculated assuming a $1.4\;
  \Msun$ pulsar.}
\end{deluxetable}

\begin{deluxetable}{lccc}                                                       
  \rotate                                                                      
  \centering                                                                    
  \tabletypesize{\scriptsize}                                                 
  \tablewidth{0pt}                                                              
  \tablecolumns{3}                                                              
  \tablecaption{Parameters of the Newly Timed Pulsars in M62 \label{table:M62_new}}
  \tablehead{
    \colhead{Parameter} &
    \colhead{J1701$-$3006D} &
    \colhead{J1701$-$3006E} &
    \colhead{J1701$-$3006F}}
  \startdata
    \cutinhead{Timing Parameters}
    Right Ascension (J2000)                       \dotfill & 17:01:13.5631(15)             & 17:01:13.2734(17)             & 17:01:12.8222(15)             \\
    Declination (J2000)                           \dotfill & $-$30:06:42.559(79)           & $-$30:06:46.89(11)            & $-$30:06:51.82(10)            \\
    Spin Period ($\s$)                            \dotfill & 0.0034177704450548(17)        & 0.0032337373331158(65)        & 0.0022947270806719(29)        \\
    Period Derivative ($\s\: \ps$)                \dotfill & $1.257(24) \times 10^{-19}$   & $3.103(27) \times 10^{-19}$   & $2.221(18) \times 10^{-19}$   \\
    Dispersion Measure ($\dmu$)                   \dotfill & 114.224(11)                   & 113.792(22)                   & 113.291(34)                   \\
    Reference Epoch (MJD)                         \dotfill & 55038.0                       & 55038.0                       & 55038.0                       \\
    Span of Timing Data (MJD)                     \dotfill & 54876--55200                  & 54876--55200                  & 54876--55200                  \\
    Number of TOAs                                \dotfill & 45                            & 40                            & 47                            \\
    RMS Timing Residual ($\us$)                   \dotfill & 6.5                           & 13.1                          & 19.5                          \\
    Error Factor                                  \dotfill & 1.1                           & 1.4                           & 1.3                           \\
    \cutinhead{Binary Parameters}
    Orbital Period (days)                         \dotfill & 1.1179034034(82)              & 0.1584774951(56)              & 0.2054870422(72)              \\
    Projected Semi-major Axis (lt-s)              \dotfill & 0.9880185(50)                 & 0.0701589(55)                 & 0.0573400(52)                 \\
    Epoch of Periastron Passage (MJD)             \dotfill & 55037.88706413(93)\tablenotemark{a} & 55037.9364630(50)       & 55038.0892403(49)             \\
    Orbital Eccentricity                          \dotfill & 0.0004122(67)                 & $< 1.9 \times 10^{-4}$        & $< 3.4 \times 10^{-4}$        \\
    Longitude of Periastron Passage (deg)         \dotfill & 101.6(1.3)                    & \nodata                       & \nodata                       \\
    \cutinhead{Derived Parameters}
    Mass Function $(\Msun)$                       \dotfill & 0.000828647(13)               & 1.47637(35)$\times$10$^{-5}$  & 4.7939(13)$\times$10$^{-6}$   \\
    Minimum Companion Mass ($\Msun$)              \dotfill & 0.12                          & 0.031                         & 0.021                         \\
    Offset from Cluster Center (core radii)       \dotfill & 0.912                         & 0.503                         & 0.185                         \\
    Intrinsic Spin-down ($\s\: \ps$)              \dotfill & $\leq 3.4 \times 10^{-19}$    & $\leq 5.56 \times 10^{-19}$   & $\leq 4.14 \times 10^{-19}$   \\
    Surface Magnetic Field ($10^9$ Gauss)         \dotfill & $\leq 1.1$                    & $\leq 1.4$                    & $\leq 0.99$                   \\
    Spin-down Luminosity ($10^{34}\; \erg\: \ps$) \dotfill & $\leq 34$                     & $\leq 65$                     & $\leq 140$                    \\
    Characteristic Age ($10^9\; \yr)$             \dotfill & $\geq 0.16$                   & $\geq 0.092$                  & $\geq 0.088$                  \\
    Mean $2\; \GHz$ Flux Density ($\uJy$)         \dotfill & 69                            & 39                            & 52                            \\
  \enddata
  \tablenotetext{a}{For M62D we give the epoch of the ascending node
    ($T\rmsub{asc}$), which is more precisely measured in the ELL1
    timing model.}  
  \tablecomments{See the notes to Table \ref{table:M62_old} for more
    details on these timing solutions.}
\end{deluxetable}

\begin{deluxetable}{lc}
  \centering
  \tabletypesize{\scriptsize}
  \tablewidth{0pt}
  \tablecolumns{2}
  \tablecaption{Parameters of NGC~6544A \label{table:NGC6544A}}
  \tablehead{
    \colhead{Parameter} &
    \colhead{J1807$-$2459A}}
  \startdata
    \cutinhead{Timing Parameters}
    Right Ascension (J2000)                       \dotfill & 18:07:20.355604(16)           \\
    Declination (J2000)                           \dotfill & $-$24:59:52.9015(65)          \\
    Spin Period ($\s$)                            \dotfill & 0.003059448798020229(30)      \\
    Period Derivative ($\s\: \ps$)                \dotfill & $-4.3352(26) \times 10^{-21}$ \\
    Dispersion Measure ($\dmu$)                   \dotfill & 134.00401(58)                 \\
    Reference Epoch (MJD)                         \dotfill & 55243.0                       \\
    Span of Timing Data (MJD)                     \dotfill & 53315--55760                  \\
    Number of TOAs                                \dotfill & 296                           \\
    RMS Timing Residual ($\us$)                   \dotfill & 1.4                           \\
    Error Factor                                  \dotfill & 1.4                           \\
    \cutinhead{Binary Parameters}
    Orbital Period (days)                         \dotfill & 0.071091483516(29)            \\
    Projected Semi-major Axis (lt-s)              \dotfill & 0.01222393(12)                \\
    Epoch of Periastron Passage (MJD)             \dotfill & 55242.99436660(24)            \\
    Orbital Eccentricity                          \dotfill & $< 1.1 \times 10^{-4}$        \\
    Longitude of Periastron Passage (deg)         \dotfill & \nodata                       \\
    Rate of Change of Orbital Period ($10^{-12}$) \dotfill & $-1.142(62) \times 10^{-12}$  \\
    \cutinhead{Derived Parameters}
    Mass Function $(\Msun)$                       \dotfill & 3.88043(11)$\times$10$^{-7}$  \\
    Minimum Companion Mass ($\Msun$)              \dotfill & 0.0092                        \\
    Offset from Cluster Center (core radii)       \dotfill & 1.31                          \\
    Intrinsic Spin-down ($\s\: \ps$)              \dotfill & $\leq 2.58 \times 10^{-19}$   \\
    Surface Magnetic Field ($10^9$ Gauss)         \dotfill & $\leq 0.9$                    \\
    Spin-down Luminosity ($10^{34}\; \erg\: \ps$) \dotfill & $\leq 36$                     \\
    Characteristic Age ($10^9\; \yr)$             \dotfill & $\geq 0.19$                   \\
    Mean $2\; \GHz$ Flux Density ($\uJy$)         \dotfill & 605                           \\
    Rotation Measure (rad$\: \pmsq$)              \dotfill & 160.4(6)                      \\
  \enddata
  \tablecomments{See the notes to Table \ref{table:M62_old} for more
    details on this timing solution.}
\end{deluxetable}

\begin{deluxetable}{lc}
  \centering
  \tabletypesize{\scriptsize}
  \tablewidth{0pt}
  \tablecolumns{2}
  \tablecaption{Parameters of NGC~6544B using DDGR Model \label{table:NGC6544B}}
  \tablehead{
    \colhead{Parameter} &
    \colhead{J1807$-$2500B}}
  \startdata
   \cutinhead{Timing Parameters}
    Right Ascension (J2000)                       \dotfill & 18:07:20.871209(53)           \\
    Declination (J2000)                           \dotfill & $-$25:00:1.915(17)            \\
    Spin Period ($\s$)                            \dotfill & 0.00418617720284089(25)       \\
    Period Derivative ($\s\: \ps$)                \dotfill & $8.23245(18) \times 10^{-20}$ \\
    Dispersion Measure ($\dmu$)                   \dotfill & 137.1535(20)                  \\
    Reference Epoch (MJD)                         \dotfill & 54881.352079                  \\
    Span of Timing Data (MJD)                     \dotfill & 53315--55760                  \\
    Number of TOAs                                \dotfill & 278                           \\
    RMS Timing Residual ($\us$)                   \dotfill & 5.4                           \\
    Error Factor                                  \dotfill & 1.0                           \\
    \cutinhead{Binary Parameters}
    Orbital Period (days)                         \dotfill & 9.9566681588(27)              \\
    Projected Semi-major Axis (lt-s)              \dotfill & 28.920391(44)                 \\
    Epoch of Periastron Passage (MJD)             \dotfill & 54881.34735776(22)            \\
    Orbital Eccentricity                          \dotfill & 0.747033198(40)               \\
    Longitude of Periastron Passage (deg)         \dotfill & 11.334600(18)                 \\
    Total System Mass ($\Msun$)                   \dotfill & 2.57190(73)                   \\
    Companion Mass ($\Msun$)                      \dotfill & 1.2064(20)                    \\
    \cutinhead{Derived Parameters}
    Pulsar Mass ($\Msun$)                         \dotfill & 1.3655(21)                    \\
    $\sin{i}$                                     \dotfill & 0.9956(14)                    \\
    Mass Function ($\Msun$)                       \dotfill & 0.2619794(12)                 \\
    Rate of Change of Orbital Period ($10^{-12}$) \dotfill & $-0.027921(62)$               \\
    Rate of Periastron Advance (deg$\, \yr^{-1}$) \dotfill & 0.0183389(35)                 \\
    Gravitational Redshift ($\s$)                 \dotfill & 0.014418(40)                  \\
    Offset from Cluster Center (core radii)       \dotfill & 4.05                          \\
    Intrinsic Spin-down ($\s\: \ps$)              \dotfill & $\leq 2.25 \times 10^{-19}$   \\
    Surface Magnetic Field ($10^9$ Gauss)         \dotfill & $\leq 0.98$                   \\
    Spin-down Luminosity ($10^{34}\; \erg\: \ps$) \dotfill & $\leq 12.0$                   \\
    Characteristic Age ($10^9\; \yr)$             \dotfill & $\geq 0.29$                   \\
    Mean $2\; \GHz$ Flux Density ($\uJy$)         \dotfill & 88                            \\
    Rotation Measure (rad$\: \pmsq$)              \dotfill & 157(1)                        \\
  \enddata
  \tablecomments{In the DDGR model, $M\rmsub{tot}$ and $M\rmsub{c}$
    are the only free PK parameters.  See the notes to Table
    \ref{table:M62_old} for more details on this timing solution.}
\end{deluxetable}

\begin{deluxetable}{lcccc}
  \rotate
  \centering
  \tabletypesize{\scriptsize}
  \tablewidth{0pt}
  \tablecolumns{5}
  \tablecaption{Parameters of the Pulsars in NGC~6624 \label{table:NGC6624}}
  \tablehead{
    \colhead{Parameter} &
    \colhead{J1823$-$3021A} &
    \colhead{J1823$-$3021B} &
    \colhead{J1823$-$3021C} &
    \colhead{J1823$-$3021D}}
  \startdata
    \cutinhead{Timing Parameters}
    Right Ascension (J2000)                       \dotfill & 18:23:40.48701(36)            & 18:23:41.5455(23)             & 18:23:41.1516(36)             & 18:23:40.5312(71)     \\
    Declination (J2000)                           \dotfill & $-$30:21:40.127(44)           & $-$30:21:40.94(47)            & $-$30:21:38.45(80)            & $-$30:21:43.66(35)    \\
    Spin Period ($\s$)                            \dotfill & 0.0054400041632727(31)        & 0.3785964921974(27)           & 0.4059359629899(53)           & 0.003020060260978(59) \\
    Period Derivative ($\s\: \ps$)                \dotfill & $3.3750(20) \times 10^{-18}$  & $2.99(14) \times 10^{-17}$    & $2.240(25) \times 10^{-16}$   & \nodata               \\
    Dispersion Measure ($\dmu$)                   \dotfill & 86.8797(56)                   & 87.01(10)                     & 86.88(24)                     & 86.8(1.1)             \\
    Reference Epoch (MJD)                         \dotfill & 55049.0                       & 55049.0                       & 55049.0                       & 55050.0               \\
    Span of Timing Data (MJD)                     \dotfill & 54898--55200                  & 54898--55200                  & 54898--55200                  & 54898--55200          \\
    Number of TOAs                                \dotfill & 33                            & 33                            & 31                            & 10                    \\
    RMS Timing Residual ($\us$)                   \dotfill & 4.6                           & 55.8                          & 104.9                         & 90.0                  \\
    Error Factor                                  \dotfill & 1.1                           & 1.0                           & 1.1                           & 3.9                   \\
    \cutinhead{Derived Parameters}
    Offset from Cluster Center (core radii)       \dotfill & 0.115                         & 3.74                          & 2.24                          & 2.33                  \\
    Intrinsic Spin-down ($\s\: \ps$)              \dotfill & $\leq 3.64 \times 10^{-16}$   & $\leq 3.45 \times 10^{-17}$   & $\leq 2.33 \times 10^{-16}$   & \nodata               \\
    Surface Magnetic Field ($10^9$ Gauss)         \dotfill & $\leq 45$                     & $\leq 120$                    & $\leq 310$                    & \nodata               \\
    Spin-down Luminosity ($10^{34}\; \erg\: \ps$) \dotfill & $\leq 8900$                   & $\leq 0.0025$                 & $\leq 0.014$                  & \nodata               \\
    Characteristic Age ($10^9\; \yr)$             \dotfill & $\geq 0.00024$                & $\geq 0.17$                   & $\geq 0.028$                  & \nodata               \\
    Mean $2\; \GHz$ Flux Density ($\uJy$)         \dotfill & 79                            & 76                            & 37                            & 29                    \\
  \enddata
  \tablecomments{See the notes to Table \ref{table:M62_old} for more
    details on these timing solutions.}
\end{deluxetable}

\end{document}